\documentclass[lettersize,journal]{IEEEtran} 
\usepackage{amsmath,amsfonts} 
\usepackage{algorithm}
\usepackage{array} %\usepackage[caption=false,font=normalsize,labelfont=sf,textfont=sf]{subfig} 
\usepackage{textcomp}
\usepackage{stfloats}
\usepackage{url}
\usepackage{amssymb}
\usepackage{verbatim}
\usepackage{graphicx}
\usepackage{subcaption} % 加载子图环境宏包 
\usepackage{cite}
\usepackage{algorithm} % 提供 algorithm 浮动环境
\usepackage{algcompatible} % 提供 algorithmic 命令（兼容旧版）
\usepackage{algpseudocode}
\usepackage{amsmath} % 提供数学符号

\usepackage{caption}
\usepackage{ragged2e} % 关键：提供允许断词的 RaggedRight

\usepackage{array}
\usepackage{makecell}
\usepackage{balance}

\usepackage[
colorlinks=true,
linkcolor=blue,
citecolor=blue,
urlcolor=magenta
]{hyperref}

\usepackage{cleveref}
\crefname{figure}{Fig.}{Figs.}

\newcommand{\StateCont}[1]{\Statex \quad\quad\quad #1}
\newcommand{\StateContm}[1]{\Statex \hspace{0.4cm} #1}
\newcommand{\StateContn}[1]{\Statex \hspace{0.6cm} #1}

\begin{document}

\title{\fontsize{24.88}{28}\selectfont Joint Beamforming Optimization and Dynamic Tracking in RIS-Enabled Secure ISAC Systems}

\author{Zhendong Li, Weichun Zhao, Zhou Su, Yan Yang, Xiaoyan Hu, \\
	Jiakang Zheng, Ying Wang, and Wen Chen

\thanks{Zhendong Li, Weichun Zhao and Xiaoyan Hu are with the School of Information and Communication Engineering, Xi’an Jiaotong University, Xi’an 710049, China (email: lizhendong@xjtu.edu.cn; zwc2856669274@stu.xjtu.edu.cn; xiaoyanhu@xjtu.edu.cn). Zhou Su is with the School of Cyber Science and Engineering, Xi'an Jiaotong University, Xi'an 710049, China (email: zhousu@ieee.org). Yan Yang is with the Rocket Force University of Engineering, Xi’an 710025, China, and also with the School of Information and Communica-tion Engineering, Xi’an Jiaotong University, Xi’an 710049, China (e-mail: yangyanyy2022@163.com). Jiakang Zheng is with the School of Electronics and Information Engineering, Beijing Jiaotong University, Beijing 100044, China (email: jiakangzheng@bjtu.edu.cn). Ying Wang is with the State Key Laboratory of Networking and Switching Technology, Beijing University of Posts and Telecommunications, Beijing 100876, China (e-mail: wangying@bupt.edu.cn). Wen Chen is with the School of Integrated Circuits, Shanghai Jiao Tong University, Shanghai 200240, China (e-mail: wenchen@sjtu.edu.cn). (Corresponding author: Zhou Su)}
\vspace{-1.5em}
}

\maketitle

\begin{abstract}
This paper investigates a reconfigurable intelligent surface (RIS)-enabled secure integrated sensing and communication (ISAC) system, where the direct links between the base station (BS) and users are blocked, and the mobile eavesdropper is treated as the sensing target. The mobility of the eavesdropper induces time-varying wiretap channels, potentially compromising secure transmission when outdated eavesdropper information is used. To address this challenge, the BS dynamically tracks the eavesdropper over consecutive time slots and exploits the predicted state information to adapt its transmission strategy. A sum secrecy rate maximization problem is formulated by jointly optimizing the BS beamforming, artificial noise, and RIS reflection coefficients, subject to a constraint on the tracking error covariance to ensure the required sensing accuracy. To tackle the resulting nonconvex optimization problem with temporal coupling, an iterative algorithm integrating an extended Kalman filter (EKF) with an alternating optimization framework is developed. Specifically, the EKF recursively predicts and updates the eavesdropper's state, while the joint optimization problem is decomposed into four tractable subproblems that are solved iteratively. Simulation results demonstrate that the proposed algorithm achieves improved secrecy performance while effectively tracking the trajectory of the mobile eavesdropper, compared with benchmark schemes.
\end{abstract}


\section{Introduction}
\IEEEPARstart{W}{ith} the evolution of mobile communication technologies, emerging applications such as the Internet-of-vehicles (IoV) and the Internet-of-things (IoT) are placing increasingly stringent demands on wireless networks, requiring not only reliable communication services but also high-precision and robust sensing capabilities. To meet these demands, integrated sensing and communication (ISAC) has emerged as a key enabling technology for sixth-generation (6G) wireless networks, enabling communication and sensing functionalities to share spectrum resources, hardware platforms, and signal processing procedures \cite{11207612}. By jointly designing communication and sensing operations, ISAC can balance the performance tradeoff between the two functionalities \cite{2}, improve spectral and energy efficiency, and reduce hardware and signal processing costs \cite{3}. Despite these advantages, the broadcast nature of wireless signals makes ISAC systems vulnerable to eavesdropping, posing significant challenges to communication confidentiality. In particular, ISAC signals used for environmental sensing may also be exploited by malicious targets to intercept confidential information embedded in the transmitted waveforms. Therefore, exploiting the sensing capability of ISAC to enhance communication security has become an important research direction.

In conventional secure communication systems, many existing schemes rely on the availability of the eavesdropper's channel state information (CSI) or at least its location \cite{8,9,10,14,19}. The sensing capability of ISAC systems offers new opportunities to enhance communication security by acquiring information about potential eavesdroppers. Accordingly, several studies have investigated secure communication in ISAC systems \cite{27,29,31}. When the exact location of a potential eavesdropper is unknown but its location distribution is available, \cite{27} exploited this probabilistic information to jointly design transmit beams and artificial noise (AN), thereby improving secrecy performance. In \cite{31}, communication beams, radar transmit beams, and receive filters were jointly optimized, with radar signals serving as interference to suppress eavesdropping. Other studies have explored resource-efficient ISAC designs under joint communication and sensing requirements. For instance, \cite{35} jointly optimized downlink beamforming, AN covariance, uplink user power allocation, and receive beamforming to minimize the total power consumption in a full-duplex secure ISAC system. In addition, spatial reconfiguration has been explored to enhance ISAC performance. Specifically, \cite{10870338} jointly optimized movable antenna positions and beamforming to maximize communication performance subject to sensing signal-to-noise ratio (SNR) constraints. Despite these advances, the performance of secure ISAC systems may deteriorate significantly in obstructed propagation environments, where severe signal attenuation and blocked propagation paths can compromise both communication reliability and sensing accuracy.

To address these challenges, reconfigurable intelligent surface (RIS) has emerged as a promising technology for enhancing wireless communication and sensing performance. The RIS typically consists of a two-dimensional metasurface comprising a large number of programmable electromagnetic elements \cite{40,9531372,8811733}. By dynamically adjusting the reflection coefficients of these elements, RIS can manipulate the propagation of electromagnetic waves and reconfigure the wireless propagation environment \cite{42,47}. This capability is particularly beneficial for secure ISAC systems, as RIS-assisted propagation paths can help overcome blocked direct links and provide additional flexibility for jointly optimizing communication and sensing performance \cite{11111722}. From a physical-layer security perspective, RIS reflection coefficients can be jointly designed with transmit beamforming and AN to enhance signal reception at legitimate users while suppressing information leakage to potential eavesdroppers \cite{11456859}. Moreover, in dynamic scenarios, adaptive RIS configuration can exploit updated channel and eavesdropper state information to accommodate time-varying propagation conditions and support secure transmission. In addition to these performance benefits, RIS offers advantages in terms of low power consumption, relatively low hardware complexity, and flexible deployment \cite{48}, making them a promising solution for enhancing secure transmission in ISAC systems.

To exploit the potential of RIS in secure ISAC systems, recent studies have investigated various RIS architectures and their associated transmission design strategies \cite{51,10979277,53,11162192,55,56}. In \cite{51}, the base station (BS) beamforming and RIS reflection coefficients were jointly optimized to maximize the secrecy rate while satisfying radar sensing performance requirements. Beyond passive RIS, active RIS have been investigated to provide additional amplification capability and potentially achieve higher beamforming gains, at the expense of increased power consumption and hardware complexity \cite{10979277}. In \cite{53}, a dual-polarized RIS was exploited to leverage the additional polarization dimension and flexible reflection control, while radar sensing signals were utilized to suppress eavesdropping and enhance communication security. Furthermore, simultaneously transmitting and reflecting RIS was shown to improve secrecy performance compared with conventional RIS configurations and RIS-free systems \cite{11162192}. These studies demonstrate the potential of diverse RIS architectures to enhance secure communication while meeting sensing requirements in ISAC systems.

Despite the advances in RIS-enabled secure ISAC, most existing studies focus on static or quasi-static scenarios, without explicitly accounting for the temporal evolution of the mobile eavesdropper's state. For example, \cite{9472958} jointly optimized RIS phase shifts, power allocation, and hybrid beamforming to maximize the sum rate in an RIS-assisted millimeter-wave non-orthogonal multiple access (NOMA) system. In \cite{11355857}, BS beamforming, RIS phase shifts, power splitting, and NOMA decoding order were jointly designed to minimize transmit power subject to communication, sensing, and energy harvesting constraints. However, these studies do not explicitly consider the dynamic tracking of mobile eavesdroppers or the impact of tracking uncertainty on secure transmission design. In the presence of the mobile eavesdropper, its time-varying position induces continuous changes in the eavesdropping channel, potentially degrading the effectiveness of transmission strategies based on outdated state information. Specifically, the intended signal may no longer be optimally directed toward legitimate users, while the AN may become less effective in suppressing eavesdropping, resulting in degraded secrecy performance. These challenges motivate the development of a dynamic secure transmission framework that integrates eavesdropper tracking with the joint optimization of BS beamforming, AN, and RIS reflection coefficients, while accounting for the required tracking accuracy.

Despite the potential of RIS to enhance communication and sensing performance and provide additional degrees of freedom for secure transmission, the security challenges posed by the mobile eavesdropper in RIS-enabled ISAC systems remain insufficiently investigated. In particular, the time-varying position of the mobile eavesdropper necessitates dynamic state estimation and adaptive transmission design to maintain secrecy performance. To address this issue, this paper adopts the eavesdropper state estimation error as a sensing performance metric and jointly optimizes the BS beamforming, AN, and RIS reflection coefficients to maximize the secrecy rate while satisfying the required tracking accuracy. However, the eavesdropper's mobility and the strong coupling among these optimization variables render the resulting problem challenging to solve. To tackle this challenge, an efficient optimization algorithm is developed. The main contributions of this paper are summarized as follows:

\begin{itemize}
	\item We propose an RIS-enabled secure ISAC system model in which the direct links between the BS and legitimate users are blocked and the eavesdropper is mobile. The RIS is partitioned into two regions for secure signal reflection and sensing echo reception, respectively, enabling enhanced signal reception at legitimate users, improved suppression of eavesdropping, and more reliable echo observations for extended Kalman filter (EKF)-based eavesdropper state estimation.
	\item We develop an efficient alternating optimization algorithm to solve the formulated dynamic optimization problem. The problem is decomposed into a sequence of per-slot optimization problems, with the EKF linking consecutive slots through eavesdropper state prediction. Within each slot, four subproblems are alternately optimized, involving BS beamforming and AN design, RIS reflection coefficient optimization for secure transmission via the alternating direction method of multipliers (ADMM), feasible solution construction for sensing-related RIS coefficients, and eavesdropping rate upper-bound updates. The resulting algorithm iteratively updates these variables until convergence.
	\item Numerical results demonstrate the convergence and effectiveness of the proposed algorithm. The algorithm converges within a small number of iterations and achieves a higher overall secrecy rate than benchmark schemes under various system configurations. Moreover, it effectively tracks the eavesdropper's trajectory over an extended time horizon, while enhancing the desired signal at legitimate users and directing AN toward the eavesdropper's spatial region to suppress information leakage.
\end{itemize}

The remainder of this paper is organized as follows. Section \ref{erbufen} presents the system model and problem formulation. Section \ref{sanbufen} develops the proposed joint beamforming optimization and dynamic tracking algorithm. Section \ref{sibufen} presents the numerical results. Finally, Section \ref{wubufen} concludes the paper.

\textit{Notations:} Scalars are denoted by lowercase letters. Vectors and matrices are respectively represented by bold lower-case and upper-case letters. ${\left(  \cdot  \right)^ * }$, ${\left(  \cdot  \right)^{\rm{T}}}$, ${\left(  \cdot  \right)^{\rm{H}}}$, ${\left(  \cdot  \right)^{ - 1}}$, ${\left[ \cdot \right]^ + }$ and ${\left[  \cdot  \right]_{m,n}}$ denote the conjugate, transpose, conjugate conjugate, inverse, positive extraction operations and $m,n$-th entry, respectively. $\left| x \right|$, $\dot x$, $\left\| {\bf{x}} \right\|$ and ${\left\| {\bf{X}} \right\|_F}$ are the magnitude of a variable $x$, the derivative of a variable $x$, the norm of a vector $\bf{x}$ and the Frobenius norm of a matrix $\bf{X}$. ${\rm{vec}}\left( {\bf{X}} \right)$ vectorizes the matrix $\bf{X}$, ${\rm{Tr}}\left( {\bf{A}} \right)$ denotes the trace of a square matrix $\bf{A}$ and ${\rm{diag}}\left( {\bf{A}} \right)$ denotes the operation of extracting the main diagonal elements of a square matrix $\bf{A}$, while $\bf{A}\succeq0 $ represents the square matrix $\bf{A}$ is a positive semidefinite matrix. ${\mathop{\rm Re}\nolimits} \left\{  \cdot  \right\}$ denotes the real parts of a complex number or vector. $ \odot $ and $ \otimes $ denote the Hadamard product and Kronecker product, respectively. $\mathbb{E}\left\{\cdot  \right\}$ is an expectation operator. $\angle b$ is the angle of complex valued $b$. ${{\bf{I}}_N}$ represents an identity matrix of size $N \times N$. Finally, the distribution of a circularly symmetric complex Gaussian (CSCG) random vector with mean $\boldsymbol{\mu} $ and covariance matrix $\bf{C}$ is given by $\mathcal{CN}\left(\boldsymbol{\mu} ,\bf{C}\right)$, and $\sim$ denotes ‘distributed as’.
\section{System Model and Problem Formulation}\label{erbufen}
\begin{figure}[!t]
	\centering
	\includegraphics[width=0.9\linewidth]{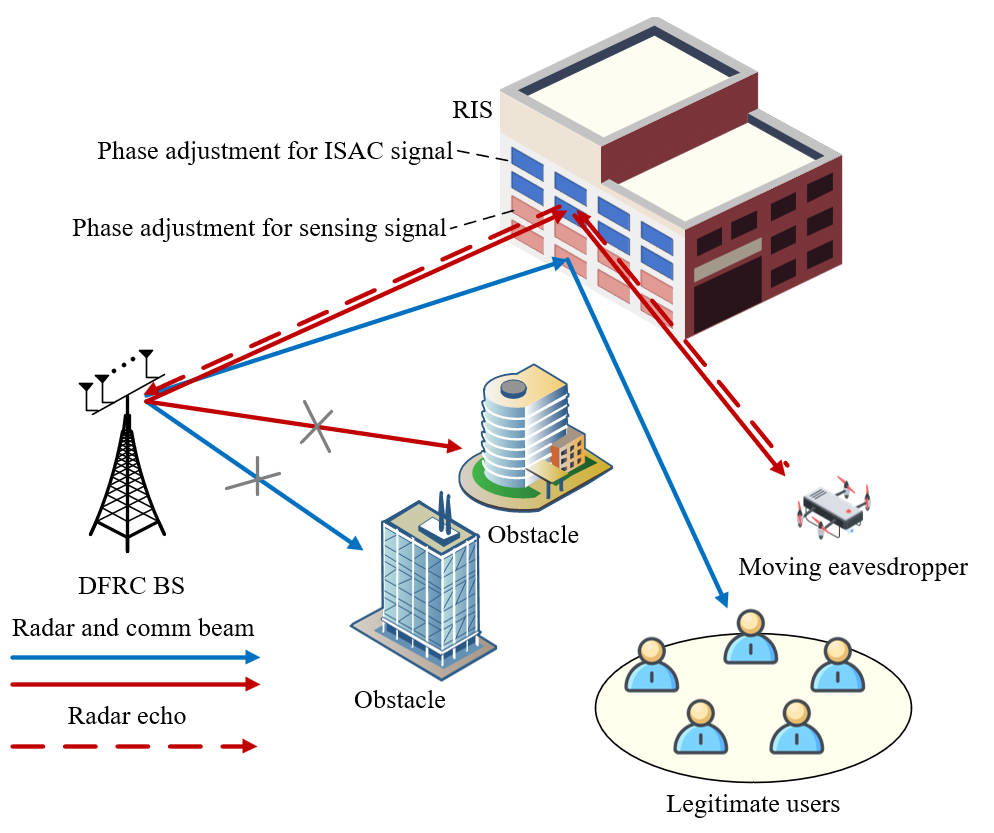}
	\caption{RIS-enabled secure ISAC systems with a moving eavesdropper.}
	\label{fig1}
\end{figure}
As shown in \cref{fig1}, this paper considers a RIS-enabled secure ISAC system. A typical application scenario is a smart industrial site, where an RIS is deployed to overcome direct-link blockage caused by surrounding buildings. Meanwhile, an unauthorized aerial vehicle (UAV) moves in three-dimensional space to intercept confidential information. Specifically, the considered system consists of a multi-antenna ISAC BS, a passive reflecting RIS, $K$ single-antenna users, and a single-antenna mobile eavesdropper\footnote{The single-eavesdropper setting is adopted to focus on the fundamental coupling between secure transmission and dynamic eavesdropper tracking. Although this work focuses on a single eavesdropper, the proposed framework can be extended to multiple non-colluding eavesdroppers by employing independent EKFs and incorporating the worst-case eavesdropping rates and individual tracking constraints into the joint optimization. A comprehensive investigation of this extension is left for future work.}. The ISAC BS is equipped with a transmit uniform linear array (ULA) comprising $N_t$ antennas and a receive ULA comprising $N_r$ antennas, while the RIS adopts a uniform planar array (UPA) with $2M$ reflecting elements. Taking path loss into account, we ignore the impact of signals propagating through scattering paths of more than two hops on the users \cite{9716123}.

This paper assumes that the BS, RIS, and users are deployed at fixed locations, whose coordinate vectors are respectively given by ${{\bf{q}}_{{\rm{bs}}}} = {\left[ {x_{\rm{bs}}},{y_{\rm{bs}}},{z_{\rm{bs}}} \right]^{\rm{T}}}$, ${{\bf{q}}_{\rm{r}}} = {\left[ {{x_{\rm{r}}},{y_{\rm{r}}},{z_{\rm{r}}}} \right]^{\rm{T}}}$, and ${{\bf{q}}_k} = {\left[ {{x_k},{y_k},{z_k}} \right]^{\rm{T}}},\forall k \in {\cal K},{\cal K} \buildrel \Delta \over = \left\{ {1,2, \ldots ,K} \right\}$, respectively. The eavesdropper moves at a low speed in the vicinity of users and wiretaps the communication between the BS and users. We analyze the system performance within a time period of duration $T$. To facilitate theoretical derivation, the total time duration $T$ is equally divided into $N$ time slots, and the duration of each time slot is $\Delta t = {T \mathord{\left/{\vphantom {T {{N}}}} \right.\kern-\nulldelimiterspace} {{N}}}$. Therefore, the location of the eavesdropper at the $n$-th time slot can be expressed as ${{\bf{q}}_{\rm{e}}}\left[ n \right] = {\left[ {{x_{\rm{e}}}\left[ n \right],{y_{\rm{e}}}\left[ n \right],{z_{\rm{e}}}\left[ n \right]} \right]^{\rm{T}}}$. Accordingly, the velocity and state vector of the eavesdropper at the $n$-th time slot are given by ${\bf{\dot q}}_{\rm e}[n]=[\dot x_{\rm e}[n],\dot y_{\rm e}[n],\dot z_{\rm e}[n]]^{\rm T}$ and $\boldsymbol{\xi}[n]=[{\bf q}_{\rm e}[n],{\bf{\dot q}}_{\rm e}[n]]^{\rm T}$. Within each time slot, the BS first transmits secure transmission signals carrying user information and AN. The signals are delivered to users with the assistance of RIS and simultaneously illuminate the mobile eavesdropper. Subsequently, the BS receives the sensing echoes reflected by the eavesdropper via the RIS. Since the signal transmission and echo reception within one time slot can be realized by fast transmit-receive switching, the self-interference caused by co-frequency simultaneous transmission and reception is neglected in this paper.
Across consecutive time slots, the system adopts the EKF algorithm to track the state of eavesdropper. Specifically, prior to the start of the service period, the BS transmits an omnidirectional waveform to search for the direction of the potential eavesdropper, and employs the combined Capon and approximate maximum likelihood method to estimate their direction and initial state \cite{10227884}. Afterwards, at the beginning of each time slot, the BS predicts the eavesdropper’s state in the current time slot based on the sensing measurements obtained in the previous time slot.

The transmit signal from the BS conveys confidential information to users on the one hand, and illuminates the eavesdropper for target sensing on the other hand. Nevertheless, the echo signals received by the BS are only utilized to estimate the state of the eavesdropper. In view of the distinct functional requirements of these two types of signals, this paper partitions the RIS into two subregions and independently regulates the secure transmission signals and sensing echo signals in each time slot. The RIS utilizes a reflection mechanism featuring tunable phase and fixed unit amplitude. We can write the reflection coefficients that correspond to the two signals as
\begin{equation}
	{\bf{\Theta }}\left[ n \right] = {\rm{diag}}\left\{ {{e^{j{\alpha _1}\left[ n \right]}}, \ldots ,{e^{j{\alpha _m}\left[ n \right]}}, \ldots ,{e^{j{\alpha _M}\left[ n \right]}}} \right\},
\end{equation}
\begin{equation}
	\!\!\!\!\!\!\!\!{\bf{\Psi }}\left[ n \right] = {\rm{diag}}\left\{ {{e^{j{\beta _1}\left[ n \right]}}, \ldots ,{e^{j{\beta _m}\left[ n \right]}}, \ldots ,{e^{j{\beta _M}\left[ n \right]}}} \right\},
\end{equation}
where ${\alpha _m}\left[ n \right] \in \left[ {0,2\pi } \right)$, $\left| {{e^{j{\alpha _m}\left[ n \right]}}} \right| = 1$, ${\beta _m}\left[ n \right] \in \left[ {0,2\pi } \right)$, $\left| {{e^{j{\beta _m}\left[ n \right]}}} \right| = 1$, $\forall m$.

\subsection{Channel Model}

The direct BS-user and BS-eavesdropper links are assumed to be blocked, and the RIS is employed to support both communication and sensing. In practice, the BS-RIS and RIS-user channels can be acquired through pilot-aided channel estimation \cite{9130088}. However, the resulting channel estimates are inevitably subject to estimation errors. In this work, perfect CSI of these links is assumed to simplify the system model and focus on the design of the proposed beamforming and tracking framework. Extending the proposed framework to imperfect CSI would require explicitly modeling channel estimation errors and incorporating their effects into the transmission design. Developing robust beamforming and tracking schemes under imperfect CSI is left for future work. Owing to the fixed deployment of the BS, RIS, and users, the BS-RIS and RIS-user channels vary relatively slowly. These channels are modeled as Rician fading channels comprising both line-of-sight (LoS) and non-line-of-sight (NLoS) components. The array response of the ULA at the BS is expressed as
	\begin{equation}
		\mathbf a_{{\rm BS},t}(\theta)
		=
		\left[
		1,\ldots,
		e^{j\frac{2\pi}{\lambda}(N_t-1)d_1\sin(\theta)}
		\right]^T
		\in\mathbb C^{N_t\times 1},
    \end{equation}
    \begin{equation}
		\mathbf a_{{\rm BS},r}(\theta)
		=
		\left[
		1,\ldots,
		e^{j\frac{2\pi}{\lambda}(N_r-1)d_1\sin(\theta)}
		\right]^T
		\in\mathbb C^{N_r\times 1},
	\end{equation}
where $\theta$ denotes the vertical angle between the BS and RIS, while $d_1$ denote the element spacings of the transmit and receive arrays at the BS. In this paper, $d_1=\lambda/2$, where $\lambda$ is the carrier wavelength. Although the RIS is divided into two regions, the expressions of its array response are actually identical. Only the array response corresponding to the region for regulating secure transmission signals is given here, which is specifically expressed as
\begin{equation}
	{{\bf{a}}_{{\rm{hor}}}}\left( {{\theta _{\rm{k}}},{\phi _{\rm{k}}}} \right) = {\left[ {1, \ldots ,\!{e^{j\left( {{M_h} -\! 1} \right)\frac{{2\pi }}{\lambda }{d_2}\sin \left( {{\phi _{\rm{k}}}} \right)\cos \left( {{\theta _{\rm{k}}}} \right)}}} \right]^{\rm{T}}},
\end{equation}
\begin{equation}
	{{\bf{a}}_{{\rm{ver}}}}\left( {{\theta _{\rm{k}}}} \right) = {\left[ {1, \ldots ,{e^{j\left( {{M_v} - 1} \right)\frac{{2\pi }}{\lambda }{d_3}\sin \left( {{\theta _{\rm{k}}}} \right)}}} \right]^{\rm{T}}},
\end{equation}
where $M=M_h M_v$, ${d_2}$ and ${d_3}$ denote the horizontal, vertical element spacings on RIS, and both are set to ${\lambda  \mathord{\left/{\vphantom {\lambda  2}} \right.\kern-\nulldelimiterspace} 2}$. ${\theta _{\rm{k}}}$ and ${\phi _{\rm{k}}}$ represent the vertical and horizontal angles from the RIS to the $k$-th user, respectively. Therefore, the overall array response of RIS can be expressed as
\begin{equation}
	{{\bf{a}}_{{\rm{UPA}}}}\left( {{\theta _{\rm{k}}},{\phi _{\rm{k}}}} \right) = {{\bf{a}}_{{\rm{hor}}}}\left( {{\theta _{\rm{k}}},{\phi _{\rm{k}}}} \right) \otimes {{\bf{a}}_{{\rm{ver}}}}\left( {{\theta _{\rm{k}}}} \right)\in\mathbb{C}^{M \times 1}.
\end{equation}

Then, the LoS channel between the BS and RIS can be expressed by
\begin{equation}
		\mathbf H_{\rm BR}^{\rm LoS}
		=
		\mathbf a_{\rm UPA}(\theta,\phi)
		\mathbf a_{{\rm BS},t}^{H}(\theta)
		\in\mathbb C^{M\times N_t},
\end{equation}
\begin{equation}
		\mathbf H_{\rm RB}^{\rm LoS}
		=
		\mathbf a_{{\rm BS},r}(\theta)
		\mathbf a_{\rm UPA}^{H}(\theta,\phi)
		\in\mathbb C^{N_r\times M}.
\end{equation}
The LoS channel between RIS and the $k$-th user is given by
\begin{equation}
	{\bf{h}}_{{k}}^{{\rm{LoS}}} = {{\bf{a}}_{{\rm{hor}}}}\left( {{\theta _k},{\phi _k}} \right) \otimes {{\bf{a}}_{{\rm{ver}}}}\left( {{\theta _k}} \right)\in\mathbb{C}^{M \times 1}.
\end{equation}
Similarly, the LoS channel between the RIS and the eavesdropper in the $n$-th time slot can be expressed as
\begin{equation}
	{\bf{h}}_{{\rm{re}}}^{{\rm{LoS}}}\left[ n \right] = {{\bf{a}}_{{\rm{hor}}}}\left( {{\theta _{\rm{e}}}\left[ n \right],{\phi _{\rm{e}}}\left[ n \right]} \right) \otimes {{\bf{a}}_{{\rm{ver}}}}\left( {{\theta _{\rm{e}}}\left[ n \right]} \right)\in\mathbb{C}^{M \times 1}.
\end{equation}

In the considered ISAC framework, the motion state of eavesdropper is recursively estimated from the received echoes using the EKF. Based on the estimated state, the RIS-eavesdropper channel is reconstructed using a geometry-based LoS model, where the estimated distance determines the large-scale attenuation, while the estimated elevation and azimuth angles specify the RIS array response. Thus, the sensing-derived state information is explicitly incorporated into the wiretap channel model and subsequently exploited for adaptive secure transmission design \cite{9966507}. Therefore, the channel gains between the BS and RIS, between the RIS and the $k$-th user and that between the RIS and eavesdropper in the $n$-th time slot can be expressed
\begin{equation}
		\!\!\mathbf H_{\rm BR}
		\!=\!
		\sqrt{
			C_0\!
			\left(\!
			\frac{d_{\rm dr}}{D_0}\!
			\right)^{-o}
		}\!
		\left(\!
		\sqrt{\!\frac{\kappa}{1\!+\!\kappa}}
		\mathbf H_{\rm BR}^{\rm LoS}
		\!+\!
		\sqrt{\!\frac{1}{1\!+\!\kappa}}
		\mathbf H_{\rm BR}^{\rm NLoS}\!
		\right),
\end{equation}
\begin{equation}
		\!\!\mathbf H_{\rm RB}
		\!=\!
		\sqrt{
			C_0\!
			\left(\!
			\frac{d_{\rm dr}}{D_0}\!
			\right)^{-o}
		}\!
		\left(\!
		\sqrt{\!\frac{\kappa}{1\!+\!\kappa}}
		\mathbf H_{\rm RB}^{\rm LoS}\!
		+\!
		\sqrt{\!\frac{1}{1\!+\!\kappa}}
		\mathbf H_{\rm RB}^{\rm NLoS}\!
		\right),
\end{equation}
\begin{equation}
\!\!\!	{{\bf{h}}_k} \!=\! \sqrt {{C_0}{{\left( {\frac{{{d_k}}}{{{D_0}}}} \right)}^{ - \rho }}}\!\! \left( {\sqrt {\frac{\vartheta }{{1 + \vartheta }}} {\bf{h}}_k^{{\rm{LoS}}} \!+\! \sqrt {\frac{1}{{1 + \vartheta }}} {\bf{h}}_k^{{\rm{NLoS}}}} \right),
\end{equation}
\begin{equation}
	{{\bf{h}}_{{\rm{re}}}}\left[ n \right] = \sqrt {{C_0}{{\left( {\frac{{{d_{{\rm{re}}}}\left[ n \right]}}{{{D_0}}}} \right)}^{ - \iota }}} {\bf{h}}_{{\rm{re}}}^{{\rm{LoS}}}\left[ n \right],
	\label{buc1}
\end{equation}
where $\left[\mathbf H_{\rm BR}^{\rm NLoS}\right]_{i,j} \sim\mathcal{CN}(0,1)$ and $\left[\mathbf H_{\rm RB}^{\rm NLoS}\right]_{i,j} \sim\mathcal{CN}(0,1)$. Here, $\kappa$ and $\vartheta$ are the Rician factors, $C_0$ is the reference path loss at $D_0=1~{\rm m}$, and $o$, $\rho$, and $\iota$ are the path loss exponents of the BS-RIS, RIS-user, and RIS-eavesdropper links, respectively.
The corresponding distances are given by $d_{{\rm{dr}}}=\left\|{\bf{q}}_{\rm r}-{\bf{q}}_{\rm bs}\right\|$, $d_k=\left\|{\bf{q}}_{\rm r}-{\bf{q}}_k\right\|$, and $d_{{\rm{re}}}[n]=\left\|{\bf{q}}_{\rm r}-{\bf{q}}_{\rm e}[n]\right\|$.

\subsection{Communication and Sensing Signal Models}

Let the transmit signal of the ISAC BS in the $n$-th time slot be expressed as
\begin{equation}
	{\bf{x}}[n]=\sum_{k=1}^{K}{\bf{w}}_k[n]s_k[n]+{\bf{z}}[n],
\end{equation}
where ${\bf{w}}_k[n]\in\mathbb{C}^{N_t \times 1}$ denotes the beamforming vector for the $k$-th user, and $s_k[n]$ is the corresponding information-bearing symbol with $\mathbb{E}\{|s_k[n]|^2\}=1$. Moreover, ${\bf{z}}[n]\in\mathbb{C}^{N_t \times 1}$ denotes the AN signal vector, which is independent of the information-bearing symbols $\{s_k[n]\}_{k=1}^{K}$. Then, the received signal at the $k$-th user in the $n$-th time slot can be written as
\begin{align}
	y_k[n] =
	& \mathbf{h}_k^{\mathrm{H}} \boldsymbol{\Theta}[n] \mathbf{H}_{\mathrm{BR}} \mathbf{w}_k[n] s_k[n] + \sum_{c\neq k}^{K} \mathbf{h}_k^{\mathrm{H}} \boldsymbol{\Theta}[n] \mathbf{H}_{\mathrm{BR}} \mathbf{w}_c[n] s_c[n] \nonumber \\
	& + \mathbf{h}_k^{\mathrm{H}} \boldsymbol{\Theta}[n] \mathbf{H}_{\mathrm{BR}} \mathbf{z}[n] + b_k,
\end{align}
where $\!{b_k} \sim {\cal C}{\cal N}\left( {0,\sigma _k^2} \right)$ denotes the additive white Gaussian noise (AWGN) introduced at the $k$-th user. The received signal at the eavesdropper is given by
\begin{align}
	y_{\mathrm{e}}[n] &= \sum_{c=1}^{K} \mathbf{h}_{\mathrm{re}}^{\mathrm{H}}[n] \boldsymbol{\Theta}[n] \mathbf{H}_{\mathrm{BR}} \mathbf{w}_c[n] s_c[n] \nonumber \\
	&\quad + \mathbf{h}_{\mathrm{re}}^{\mathrm{H}}[n] \boldsymbol{\Theta}[n] \mathbf{H}_{\mathrm{BR}} \mathbf{z}[n] + b_{\mathrm{e}}. \label{received_signal_eavesdropper}
\end{align}
where ${b_{\rm{e}}} \sim {\cal C}{\cal N}\left( {0,\sigma _{\rm{e}}^2} \right)$ denotes the AWGN introduced at the eavesdropper. The direct BS-RIS-BS reflection component is mainly determined by the static BS-RIS link and does not contain the eavesdropper's motion information. Hence, it can be calibrated and suppressed by standard clutter cancellation methods. After removing this quasi-static component, the residual echo associated with the eavesdropper is given by
\begin{equation}
	\mathbf y_r[n]
	=
	\sum_{c=1}^{K}
	\mathbf H[n]\mathbf w_c[n]s_c[n]
	+
	\mathbf H[n]\mathbf z[n]
	+
	\mathbf b[n],
	\label{received_echo_signal}
\end{equation}
where $\mathbf H[n]=\mathbf H_{\rm RB}\boldsymbol{\Psi}[n]\mathbf h_{\rm re}[n]\mathbf h_{\rm re}^{\rm H}[n]\boldsymbol{\Theta}[n]\mathbf H_{\rm BR}$, and $\mathbf{b}[n]\sim\mathcal{CN}\!\left(\mathbf 0,\sigma_r^2\mathbf I_{N_r}\right)$ denotes the AWGN introduced at the BS. Since both RIS subregions contribute to the cascaded sensing channel in \eqref{received_echo_signal}, fixed equal-size partition is used as the baseline. Dynamically adjusting the number of elements for the two subregions offers extra flexibility to balance secrecy performance and tracking accuracy. Nevertheless, this design introduces additional discrete optimization variables coupled with BS beamforming and RIS reflection coefficients, which increases the complexity of the joint optimization problem.

\subsection{Secure Communication and Sensing Performance Metrics}

\subsubsection{Secure Communication Metrics}

We focus on the total secrecy rate between the BS and all users. The secrecy performance is determined by the achievable communication rate of each user and the corresponding information leakage rate at the eavesdropper. According to the Shannon formula, the achievable communication rate between the BS and the $k$-th user in the $n$-th time slot can be expressed as
\begin{equation}
	R_k[n]
	=
	\log_2
	\left(
	1+
	\frac{
		\left|
		\mathbf h_k^{\rm H}
		\boldsymbol{\Theta}[n]
		\mathbf H_{\rm BR}
		\mathbf w_k[n]
		\right|^2
	}{
		P_k[n]
	}
	\right),
	\label{user_rate}
\end{equation}
where $P_k[n]
\!\!=\!\!\!
\sum_{\substack{c\neq k}}^{K}\!\!
\left|
\mathbf h_k^{\rm H}\!
\boldsymbol{\Theta}[n]
\mathbf H_{\rm BR}\!
\mathbf w_c[n]
\right|^2
\!\!+\!
\left|\!
\mathbf h_k^{\rm H}\!
\boldsymbol{\Theta}[n]
\mathbf H_{\rm BR}
\mathbf z[n]
\right|^2
\!+\!
\sigma_k^2$. To obtain a conservative secrecy rate, we adopt the worst case eavesdropping model in \cite{10054167}, where the eavesdropper employs ideal successive interference cancellation and decodes the $k$-th user's signal after canceling the other users' signals. This gives an upper bound on the eavesdropping rate and a lower bound on the secrecy rate. Therefore, the eavesdropping rate of the $k$-th user can be expressed as
\begin{equation}
	C_k[n]
	=
	\log_2
	\left(
	1+
	\frac{
		\left|
		\mathbf h_{\rm re}^{\rm H}[n]
		\boldsymbol{\Theta}[n]
		\mathbf H_{\rm BR}
		\mathbf w_k[n]
		\right|^2
	}{
		\left|
		\mathbf h_{\rm re}^{\rm H}[n]
		\boldsymbol{\Theta}[n]
		\mathbf H_{\rm BR}
		\mathbf z[n]
		\right|^2
		+
		\sigma_e^2
	}
	\right).
	\label{eavesdropping_rate}
\end{equation}
Therefore, the achievable secrecy rate between the BS and the $k$-th user can be expressed as
\begin{equation}
	{D_k}\left[ n \right] = {\left[ {{R_k}\left[ n \right] - {C_k}\left[ n \right]} \right]^ + }.
\end{equation}

\subsubsection{Sensing Metrics}

The sensing performance is characterized by the tracking accuracy of the mobile eavesdropper. With the aid of the RIS, the eavesdropper-related measurements can be extracted from echo at the BS. After matched filtering, the BS obtains the observation parameters $\hat \tau[n]$, $\hat \mu[n]$, $\hat \theta_{\rm e}[n]$, and $\hat \phi_{\rm e}[n]$, which correspond to the round-trip time delay, Doppler shift, vertical angle, and horizontal angle, respectively. Their relationship with the eavesdropper state $\boldsymbol{\xi}[n]$ is modeled as
\begin{equation}
	\label{canshu1}
	\hat \tau \left[ n \right] = \frac{{2{d_{{\rm{re}}}}\left[ n \right]}}{c} + {w_{\tau \left[ n \right]}},
\end{equation}
\begin{equation}
	\label{canshu2}
	\hat \mu \left[ n \right] = \frac{{2{f_c}{\bf{\dot q}}_{\rm{e}}^{\rm{T}}\left[ n \right]\left( {{{\bf{q}}_{\rm{e}}}\left[ n \right] - {{\bf{q}}_{\rm{r}}}} \right)}}{{c{d_{{\rm{re}}}}\left[ n \right]}} + {w_{\mu \left[ n \right]}},
\end{equation}
\begin{equation}
	\label{canshu3}
	\sin \hat \theta_{\rm{e}} \left[ n \right] = \frac{{{z_{\rm{r}}} - {z_{\rm{e}}}\left[ n \right]}}{{{d_{{\rm{re}}}}\left[ n \right]}} + {w_{\sin \theta_{\rm{e}} \left[ n \right]}},
\end{equation}
\begin{equation}
	\label{canshu4}
	\cos \hat \theta_{\rm{e}} \left[ n \right]\! =\! \frac{{\sqrt {{{\left( {{x_{\rm{e}}}\left[ n \right] \!-\! {x_{\rm{r}}}} \right)}^2} + {{\left( {{y_{\rm{e}}}\left[ n \right] \!-\! {y_{\rm{r}}}} \right)}^2}} }}{{{d_{{\rm{re}}}}\left[ n \right]}} \!+\! {w_{\cos \theta_{\rm{e}} \left[ n \right]}},
\end{equation}
\begin{equation}
	\label{canshu5}
	\sin \hat \phi_{\rm{e}} \left[ n \right] \!=\! \frac{{{y_{\rm{e}}}\left[ n \right] - {y_{\rm{r}}}}}{{\sqrt {{{\left( {{x_{\rm{e}}}\left[ n \right]\! -\! {x_{\rm{r}}}} \right)}^2} + {{\left( {{y_{\rm{e}}}\left[ n \right]\! -\! {y_{\rm{r}}}} \right)}^2}} }} + {w_{\sin \phi_{\rm{e}} \left[ n \right]}},
\end{equation}
where ${f_c}$ and $c$ denote the carrier frequency and the speed of light, respectively. The terms ${w_{\tau \left[ n \right]}}$, ${w_{\mu \left[ n \right]}}$, ${w_{\sin \theta_{\rm{e}} \left[ n \right]}}$, ${w_{\cos \theta_{\rm{e}} \left[ n \right]}}$, and ${w_{\sin \phi_{\rm{e}} \left[ n \right]}}$ denote zero-mean Gaussian measurement noises with variances $\sigma _{\tau \left[ n \right]}^2$, $\sigma _{\mu \left[ n \right]}^2$, $\sigma _{\sin \theta_{\rm{e}} \left[ n \right]}^2$, $\sigma _{\cos \theta_{\rm{e}} \left[ n \right]}^2$, and $\sigma _{\sin \phi_{\rm{e}} \left[ n \right]}^2$, respectively. When $\sigma _{\theta_{\rm{e}} \left[ n \right]}^2$ and $\sigma _{\phi_{\rm{e}} \left[ n \right]}^2$ are sufficiently small, we can approximately assume that $\sigma _{\sin \theta_{\rm{e}} \left[ n \right]}^2 \approx \sin \hat \theta_{\rm{e}} \left[ n \right]\sigma _{\theta_{\rm{e}} \left[ n \right]}^2$, $\sigma _{\cos \theta_{\rm{e}} \left[ n \right]}^2 \approx \cos \hat \theta_{\rm{e}} \left[ n \right]\sigma _{\theta_{\rm{e}} \left[ n \right]}^2$, and $\sigma _{\sin \phi_{\rm{e}} \left[ n \right]}^2 \approx \sin \hat \phi_{\rm{e}} \left[ n \right]\sigma _{ \phi_{\rm{e}} \left[ n \right]}^2$. Since the measurement noise is inversely proportional to the SNR of the received echo, we have $\sigma _{\tau \left[ n \right]}^2 = {{{a_1}} \mathord{\left/{\vphantom {{{a_1}}{{\rm{SNR}}}}} \right.\kern-\nulldelimiterspace} {{\rm{SNR}}}}$, $\sigma _{\mu \left[ n \right]}^2 = {{{a_2}} \mathord{\left/{\vphantom {{{a_2}} {{\rm{SNR}}}}} \right.\kern-\nulldelimiterspace} {{\rm{SNR}}}}$, $\sigma _{\theta \left[ n \right]}^2 = {{{a_3}} \mathord{\left/{\vphantom {{{a_3}} {{\rm{SNR}}}}} \right.\kern-\nulldelimiterspace} {{\rm{SNR}}}}$, and $\sigma _{\phi \left[ n \right]}^2 = {{{a_4}} \mathord{\left/{\vphantom {{{a_4}} {{\rm{SNR}}}}} \right.\kern-\nulldelimiterspace} {{\rm{SNR}}}}$. According to \cite{68}, ${a_1}$, ${a_2}$, ${a_3}$, and ${a_4}$ are constants determined by the system structure and the adopted signal processing method. The \rm{SNR} can be expressed as
\begin{equation}
	\label{11}
	{\rm SNR}
	=
	\frac{
		\left\|
		\mathbf H[n]\mathbf x[n]
		\right\|^2
	}{
		\sigma_r^2
	}.
\end{equation}

Let ${\bf{\Omega }}\left[ n \right] = {\left[ {\hat \tau \left[ n \right],\hat \mu \left[ n \right],\sin \hat \theta_{\rm{e}} \left[ n \right],\cos \hat \theta_{\rm{e}} \left[ n \right],\sin \hat \phi_{\rm{e}} \left[ n \right]} \right]^{\rm{T}}}$. Therefore, \eqref{canshu1}--\eqref{canshu5} can be rewritten as
\begin{equation}
	\label{celiangmoxing}
	{\bf{\Omega }}\left[ n \right] = {g_{{n}}}\left( {\boldsymbol{\xi}[n]} \right) + {\bf{m}}\left[ n \right],
\end{equation}
where ${\bf{m}}\!\left[ n \right]\!\! =\!\! {\left[ {{w_{\tau \left[ n \right]}},{w_{\mu \left[ n \right]}},{w_{\sin \theta_{\rm{e}} \left[ n \right]}},{w_{\cos \theta_{\rm{e}} \left[ n \right]}},{w_{\sin \phi_{\rm{e}} \left[ n \right]}}} \right]^{\!\rm{T}}}$ denotes the zero-mean Gaussian noise vector. The covariance matrix of ${\bf{m}}\left[ n \right]$ is denoted as ${{\bf{Q}}_{{\bf{m}}\left[ n \right]}}$, which is given by
\begin{align}
	\label{tag1}
	\mathbf{Q}_{\mathbf{m}[n]}
	&= \rm{diag}\Bigg(
	\frac{a_1}{\rm{SNR}},\,
	\frac{a_2}{\rm{SNR}},\,
	\frac{a_3\sin^2\!\big(\hat{\theta}_{\rm e}[n]\big)}{\rm{SNR}}, \nonumber\\
	&\qquad\qquad
	\frac{a_3\cos^2\!\big(\hat{\theta}_{\rm e}[n]\big)}{\rm{SNR}},\,
	\frac{a_4\sin^2\!\big(\hat{\phi}_{\rm e}[n]\big)}{\rm{SNR}}
	\Bigg).
\end{align}

Since the eavesdropper is assumed to perform random motion, the state transition model can be expressed as
\begin{equation}
	\boldsymbol{\xi}\left[ {n + 1} \right] = {{\bf{S}}_{{\rm{cv}}}}\boldsymbol{\xi}[n] + {{\bf{z}}_{\rm{t}}}\left[ n \right],
	\label{buc2}
\end{equation}
where ${{\bf{S}}_{{\rm{cv}}}}$ denotes the state transition matrix, and ${{\bf{z}}_{\rm{t}}}\left[ n \right]$ is the process noise vector. ${{\bf{z}}_{\rm{t}}}\left[ n \right]$ characterizes the mismatch between the adopted model and the actual moving behavior of the eavesdropper. Its covariance matrix is expressed as
\begin{equation}
	{{\bf{Q}}_{\boldsymbol{\xi}}}\! =\! {\rm{diag}}\!\left( {\frac{1}{4}{\delta ^4}\sigma _x^2,\frac{1}{4}{\delta ^4}\sigma _y^2,\frac{1}{4}{\delta ^4}\sigma _z^2,{\delta ^2}\sigma _{\dot x}^2,{\delta ^2}\sigma _{\dot y}^2,{\delta ^2}\sigma _{\dot z}^2} \right),
	\label{buc3}
\end{equation}
where ${\left[ {\sigma _x^2,\sigma _y^2,\sigma _z^2} \right]^{\rm{T}}}$ and ${\left[ {\sigma _{\dot x}^2,\sigma _{\dot y}^2,\sigma _{\dot z}^2} \right]^{\rm{T}}}$ represent the Gaussian noise variances of the position and velocity components along the $x$, $y$, and $z$ axes, respectively. The entries of ${\bf{Q}}_{\boldsymbol{\xi}}$  quantify the process uncertainties in the position and velocity evolution along the three spatial axes. Since the measurement model in \eqref{celiangmoxing} is nonlinear, the EKF is adopted to linearize it. The Jacobian matrix ${{\partial {g_{{n}}}} \mathord{\left/{\vphantom {{\partial {g_{\rm{n}}}} {\partial {\bf{\xi }}\left[ n \right]}}} \right.\kern-\nulldelimiterspace} {\partial \boldsymbol{\xi}[n]}}$ is denoted by ${{\bf{G}}_{{n}}}$.

Based on the sensing measurements obtained in the previous time slot, the EKF predicts and updates the state of the eavesdropper in the current time slot. The detailed procedure is given as follows.

\text{(i)} \textit{Prediction Part}:
\begin{equation}
	{\bf{\hat {\boldsymbol{\xi}} }}\left[ {n\left| {n - 1} \right.} \right] = {{\bf{S}}_{\rm{cv}}}{\boldsymbol{\xi}}\left[ {n - 1} \right],
\end{equation}

\text{(ii)} \textit{Predicted Mean Squared Error (MSE) Matrix}:
\begin{equation}
	{\bf{M}}\left[ {n\left| {n - 1} \right.} \right] = {{\bf{S}}_{\rm{cv}}}{\bf{M}}\left[ {n - 1} \right]{\bf{S}}_{\rm{cv}}^{\rm{H}} + {{\bf{Q}}_{\boldsymbol{\xi}}},
\end{equation}

\text{(iii)} \textit{Kalman Gain Matrix}:
\begin{equation}
	\label{kaermanzengyi}
	{\bf{K}}\left[ n \right]\! =\! {\bf{M}}\!\left[ {n\!\left| {n\! -\! 1}\! \right.} \right]\!{\bf{G}}_{{n}}^{\rm{H}}{\left( {{{\bf{Q}}_{{\bf{m}}\left[ n \right]}}\! +\! {{\bf{G}}_{{n}}}{\bf{M}}\!\left[ {n\!\left| {n\! -\! 1}\! \right.} \right]\!{\bf{G}}_{{n}}^{\rm{H}}} \right)^{ - 1}},
\end{equation}

\text{(iv)} \textit{Correction Part}:
\begin{equation}
	{\bf{\hat {\boldsymbol{\xi}} }}\left[ n \right]\! =\! {\bf{\hat {\boldsymbol{\xi}} }}\left[ {n\!\left| {n\! -\! 1} \right.} \right] + {\bf{K}}\left[ n \right]\left( {{\bf{\Omega }}\left[ n \right] - {g_{{n}}}\left( {{\bf{\hat {\boldsymbol{\xi}} }}\left[ {n\left| {n - 1} \right.} \right]} \right)} \right),
\end{equation}

\text{(v)} \textit{Updated MSE Matrix}:
\begin{equation}
	\label{gengxinjunfangwuchajuzhen}
	{\bf{M}}\left[ n \right] = \left( {{\bf{I}} - {\bf{K}}\left[ n \right]{{\bf{G}}_{{n}}}} \right){\bf{M}}\left[ {n\left| {n - 1} \right.} \right].
\end{equation}

The updated MSE matrix $\mathbf{M}[n]$ in \eqref{gengxinjunfangwuchajuzhen} is used to evaluate the tracking accuracy of the mobile eavesdropper. In the considered dynamic tracking scenario, the estimated velocity is used to predict the eavesdropper's position in subsequent time slots, and velocity estimation errors can consequently affect future position prediction accuracy. Therefore, $\mathrm{Tr}(\mathbf{M}[n])$ is adopted to characterize the joint estimation uncertainty of position and velocity, thereby supporting reliable continuous tracking of the mobile eavesdropper. As shown in \eqref {11} and \eqref {tag1}, the measurement noise covariance depends on the received echo SNR, which is jointly affected by the BS beamforming vectors, the AN signal, and the RIS reflection coefficients. Therefore, the sensing accuracy is inherently coupled with the secure communication design. Since the RIS-eavesdropper channel in \eqref{buc1} is reconstructed from EKF-predicted states, position prediction errors yield deviations in distance-dependent path loss and RIS array response, which impair the optimized beamforming and AN. While the process noise in \eqref{buc2} and \eqref{buc3} captures stochastic mismatches with the motion model, EKF alleviates such errors via recursive measurement updates. However, severe or persistent motion-model mismatch still causes inaccurate channel reconstruction and deteriorated secrecy performance.
\subsection{Problem Formulation}

We aim to maximize the secrecy rate between the BS and all users while guaranteeing the sensing accuracy of the mobile eavesdropper. Specifically, in each time slot, the BS designs the beamforming vectors $\{{\bf{w}}_k[n]\}$ and the AN vector ${\bf{z}}[n]$ based on the predicted eavesdropper state. Meanwhile, the RIS reflection coefficient matrices ${\bf{\Theta }}[n]$ and ${\bf{\Psi }}[n]$ are optimized to regulate the secure transmission signal and the echo sensing signal, respectively. Thus, the $n$-th time slot optimization problem is formulated as problem \text{P1}
\begin{subequations}
	\begin{align}
		\text{P1:} \quad &\max_{\{\mathbf{w}_k[n]\}, \mathbf{z}[n], \mathbf{\Theta}[n], \mathbf{\Psi}[n] }
		\sum\limits_{k = 1}^K {D_k[n]},\nonumber \\
		&\quad\quad\quad\quad \text{s.t.} \quad 
		\mathrm{Tr}(\mathbf{M}[n]) \le \Gamma_{\max},\label{p1a}\\
		& \quad \quad \quad\quad\quad\quad
		\sum\limits_{i=1}^{K}\! \|\mathbf{w}_i[n]\|^2 \!\!+\!\! \|\mathbf{z}[n]\|^2 \!\!\le\! P_{\max},\\
		& \quad \quad \quad\quad\quad\quad
		\left| \left[\mathbf{\Theta}[n]\right]_{m,m} \right| = 1,\label{p1d}\\
		& \quad \quad \quad\quad\quad\quad
		\left| \left[\mathbf{\Psi}[n]\right]_{m,m} \right| = 1,\label{p1e}
	\end{align}
\end{subequations}
where ${\Gamma _{\max }}$ denotes the maximum tolerable tracking MSE, and ${P_{\max }}$ is the transmit power budget of the BS. Constraint \eqref{p1a} ensures the EKF-based tracking accuracy, while constraints \eqref{p1d} and \eqref{p1e} enforce the unit-modulus RIS reflection coefficients for secure transmission and echo sensing, respectively. Due to the coupled variables, the fractional objective, and the non-convex constraints \eqref{p1a}, \eqref{p1d}, and \eqref{p1e}, problem \text{P1} is difficult to solve directly. Therefore, an alternating optimization algorithm is developed to solve it.

\section{Joint Beamforming Optimization and Dynamic Tracking Algorithm Design}\label{sanbufen}

In this section, we develop an alternating optimization algorithm to solve problem \text{P1}. Specifically, we first reformulate the original problem by introducing auxiliary variables, and then decompose the transformed problem into four tractable subproblems that are solved alternately.

\subsection{Problem Transformation}

To facilitate the reformulation of the secrecy rate maximization problem, we introduce auxiliary variables $\{A_k\}$ to upper bound the eavesdropping rates. Thus, the eavesdropping rate term can be separated from the legitimate rate term and handled by an additional constraint. Defining $\mathbf H_k[n]
=
\mathbf h_k^{\rm H}
\boldsymbol{\Theta}[n]
\mathbf H_{\rm BR}
$ and $\mathbf H_e[n]
=
\mathbf h_{\rm re}^{\rm H}[n]
\boldsymbol{\Theta}[n]
\mathbf H_{\rm BR}$, the objective function of problem \text{P1} can be rewritten as
\begin{equation}
	\!\sum\limits_{k = 1}^K \!\!{\left[\! {{{\log }_2}\!\!\left(\!\!\! {1\! \!+ \frac{{{{\left| {{{\bf{H}}_k}\left[ n \right]{{\bf{w}}_k}\left[ n \right]} \right|}^2}}}{\!\!\!\!{\sum\limits_{c \ne k}^K \!{{{\left| {{{\bf{H}}_k}\!\left[ n \right]\!{{\bf{w}}_c}\!\left[ n \right]} \right|}^2}} \!\! + \!\!{{\left| {{{\bf{H}}_k}\!\left[ n \right]\!{\bf{z}}\!\left[ n \right]} \right|\!}^2}\!\!\! +\! \sigma _k^2}}} \right)\! \!\!-\!\! {A_k}} \right]}\!.
\end{equation}
The eavesdropping rate upper bound constraint is given by
\begin{equation}
	{C_k}\left[ n \right] \le {A_k},\quad A_k\geq 0,\quad \forall k. \label{p1b}
\end{equation}

Since a smaller $A_k$ leads to a larger objective value, the optimal solution satisfies the boundary condition of \eqref{p1b}, namely, $A_k^\star=C_k[n]$. According to the definition of $C_k[n]$, constraint \eqref{p1b} can be equivalently transformed into
\begin{equation}
	\label{qietingyueshu}
	\!{\left| {{{\bf{H}}_{\rm{e}}}\!\left[ n \right]{{\bf{w}}_k}\!\left[ n \right]} \right|^2}\! +\! \left( {1\! -\! {2^{{A _k}}}} \right){\left| {{{\bf{H}}_{\rm{e}}}\!\left[ n \right]{\bf{z}}\!\left[ n \right]} \right|^2} \le \left( {{2^{{A _k}}}\! -\! 1} \right)\sigma _{\rm{e}}^2.
\end{equation}

Next, to handle the fractional structure in the legitimate communication rate, we apply the Lagrangian dual reformulation and quadratic transform by introducing auxiliary variables ${\bf{r}} \buildrel \Delta \over = {\left[ {{r_1},{r_2}, \ldots ,{r_K}} \right]^{\rm{T}}}$ and ${\bf{c}}\! \buildrel \Delta \over =\! {\left[ {{c_1},{c_2}, \ldots ,{c_K}} \right]^{\rm{T}}}$. Here, $r_k$ is associated with the SINR of the $k$-th user, while $c_k$ is used to decouple the desired signal term from the interference-plus-noise term. Thus, the objective function is transformed into
\begin{align}
	&\sum\limits_{k = 1}^K \!{{{\log }_2}\left( {1\! +\! {r_k}} \right)} \! -\! \sum\limits_{k = 1}^K\! {{r_k}} \! +\! \sum\limits_{k = 1}^K \!{2\sqrt {1 \!+\! {r_k}} {\mathop{\rm Re}\nolimits}\! \left\{ {c_k^ * {{\bf{H}}_k}\left[ n \right]{{\bf{w}}_k}\left[ n \right]} \right\}} \nonumber\\
	& - \sum\limits_{k = 1}^K {{{\left| {{c_k}} \right|}^2}\sigma _k^2}  - \sum\limits_{k = 1}^K {{{\left| {{c_k}} \right|}^2}\sum\limits_{c = 1}^K {{{\left| {{{\bf{H}}_k}\left[ n \right]{{\bf{w}}_c}\left[ n \right]} \right|}^2}} }  - \sum\limits_{k = 1}^K {{A_k}}  \nonumber \\
	&- \sum\limits_{k = 1}^K {{{\left| {{c_k}} \right|}^2}{{\left| {{{\bf{H}}_k}\left[ n \right]{\bf{z}}\left[ n \right]} \right|}^2}}.
\end{align}

To facilitate the subsequent derivations, we rewrite the objective function in an explicit and compact form by introducing the following equivalent notations. Specifically, define ${\bf f}[n]=\mathrm{diag}\!\left({\boldsymbol{\Theta}}[n]\right)$, let ${\bf W}[n]\!=\!\big[{\bf w}_1[n],{\bf w}_2[n],\ldots,{\bf w}_K[n]\big]$ and denote the stacked beamforming vector as ${\bf w}[n]\!=\!\mathrm{vec}\!\left({\bf W}[n]\right)$. With these definitions, the objective function can be given by
\begin{align}
	F &=\! {\mathop{\rm Re}\nolimits}\! \left\{ {{{\bf{g}}^{\rm{H}}}{\bf{f}}\left[ n \right]} \right\} \!-\! {{\bf{f}}^{\rm{H}}}\left[ n \right]{\bf{Df}}\left[ n \right] \!+\! {\varepsilon _1}, \nonumber\\
	&=\! {\mathop{\rm Re}\nolimits}\! \left\{\! {{{\bf{a}}^{\rm{H}}}{\bf{w}}\!\left[ n \right]}\! \right\} - {\left\| {{{\bf{B}}_1}\!{\bf{w}}\!\left[ n \right]}\! \right\|^2} -\!\! {\left\| {{{\bf{B}}_2}{\bf{z}}\!\left[ n \right]}\! \right\|^2} \!+\! {\varepsilon _1}.
\end{align}

\textit{Proof:} Please refer to Appendix A.$\hfill\blacksquare$

Therefore, problem \text{P1} can be transformed into problem \text{P2}, which is given by
\begin{subequations}
	\begin{align}
		\text{P2:}~\operatorname*{max}_{\substack{
				\mathbf{w}[n],\, \mathbf{z}[n],\, \mathbf{f}[n],
				\mathbf{\Psi}[n],\,\\ \{A_k\},\, \mathbf{r},\, \mathbf{c}
		}} \  &F, \nonumber\\
		\text{s.t.}\quad\quad\quad\quad
		&\eqref{p1a}, \eqref{qietingyueshu}, \eqref{p1e}, \nonumber \\
		& A_k\geq 0,\quad \forall k, \nonumber\\
		&{\left\| {{\bf{w}}\left[ n \right]} \right\|^2} + {\left\| {{\bf{z}}\left[ n \right]} \right\|^2} \!\le\! {P_{\max }},\label{p2c}\\
		& \left| {{{\left[ {{\bf{f}}\left[ n \right]} \right]}_m}} \right| = 1. \label{p2d}
	\end{align}
\end{subequations}
Given the other variables, the transformed objective function $F$ is concave with respect to each individual block, which enables a block-wise iterative solution. However, problem \text{P2} remains non-convex due to the EKF-based tracking constraint \eqref{p1a}, the eavesdropping rate constraint \eqref{qietingyueshu}, and the unit-modulus constraints \eqref{p1e} and \eqref{p2d}. Therefore, we solve \text{P2} by alternating optimization in the following subsection.

\subsection{Problem Solution}

To solve problem \text{P2}, we adopt an alternating optimization framework. Specifically, the auxiliary variables $\mathbf r$ and $\mathbf c$, the BS beamforming variables, and the RIS reflection coefficients are optimized in an alternating manner. After each round of block updates, $\{A_k\}$ is updated based on the current eavesdropping rates. The detailed update procedures are given below. For given beamforming and RIS reflection coefficients, the optimal auxiliary variable $r_k$ can be obtained as
\begin{equation}
	\label{gengxin1}
	r_k^{ \star \,} = \frac{{{{\left| {{{\bf{H}}_k}\left[ n \right]{{\bf{w}}_k}\left[ n \right]} \right|}^2}}}{{\sum\limits_{c \ne k}^K {{{\left| {{{\bf{H}}_k}\left[ n \right]{{\bf{w}}_c}\left[ n \right]} \right|}^2}}  \!+\! {{\left| {{{\bf{H}}_k}\left[ n \right]{\bf{z}}\left[ n \right]} \right|}^2} \!+\! \sigma _k^2}}.
\end{equation}
Similarly, the optimal solution of auxiliary variable ${\bf{c}}$ can be obtained as
\begin{equation}
	\label{gengxin2}
	c_k^ \star  = \frac{{\sqrt {1 + {r_k}} {{\bf{H}}_k}\left[ n \right]{{\bf{w}}_k}\left[ n \right]}}{{\sum\limits_{c = 1}^K {{{\left| {{{\bf{H}}_k}\left[ n \right]{{\bf{w}}_c}\left[ n \right]} \right|}^2}}  + {{\left| {{{\bf{H}}_k}\left[ n \right]{\bf{z}}\left[ n \right]} \right|}^2} + \sigma _k^2}}.
\end{equation}

After updating $\mathbf r$ and $\mathbf c$, the remaining variables are optimized block by block.

\subsubsection{Active Beamforming for ISAC BS}

For subproblem 1, given the RIS reflection coefficients ${\bf{\Psi }}\left[ n \right]$ for the echo sensing signal, the reflection coefficients ${\bf{f}}\left[ n \right]$ for the secure transmission signal, and the eavesdropping rate upper bound $A_k$, problem \text{P2} can be transformed into problem \text{P3}, which is expressed as
\begin{subequations}
	\begin{align}
		\text{P3:}~\operatorname*{max}_{\mathbf{w}[n], \mathbf{z}[n]} &F, \nonumber \\
		\text{s.t.}\quad
		&\eqref{p1a}, \eqref{qietingyueshu},\eqref{p2c}. \nonumber 
	\end{align}
\end{subequations}
Next, we handle the non-convex constraints in \text{P3}. For constraint \eqref{p1a}, by substituting \eqref{kaermanzengyi} into \eqref{gengxinjunfangwuchajuzhen}, the inverse matrix of ${\bf{M}}\left[ n \right]$ can be obtained as
\setcounter{equation}{46}
\begin{equation}
	{{\bf{M}}^{ - 1}}\left[ n \right] = {{\bf{M}}^{ - 1}}\left[ {n\left| {n - 1} \right.} \right] + {\bf{G}}_{{n}}^{\rm{H}}{\bf{Q}}_{{\bf{m}}\left[ n \right]}^{ - 1}{{\bf{G}}_{{n}}},
\end{equation}
where ${\bf{M}}\left[ {n\left| {n - 1} \right.} \right]$ denotes the predicted MSE matrix. Based on \eqref{tag1}, we rewrite ${\bf{Q}}_{{\bf{m}}\left[ n \right]}^{ - 1}$ as ${\bf{Q}}_{{\bf{m}}\left[ n \right]}^{ - 1} = {\rm{SNR}} \cdot {\bf{\tilde Q}}_{{\bf{m}}\left[ n \right]}^{ - 1}$, where ${\rm{SNR}} \!\!=\!\! {{\left( {\sum\nolimits_{k = 1}^K \!{{{\left\| {{\bf{H}}\!\left[ n \right]{{\bf{w}}_k}\left[ n \right]} \right\|}^2}}  + {{\left\| {{\bf{H}}\left[ n \right]{\bf{z}}\left[ n \right]} \right\|}^2}} \right)} \mathord{\left/{\vphantom {{\left( {\sum\nolimits_{k = 1}^K {{{\left| {{\bf{H}}\left[ n \right]{{\bf{w}}_k}\left[ n \right]} \right|}^2}}  + {{\left| {{\bf{H}}\left[ n \right]{\bf{z}}\left[ n \right]} \right|}^2}} \right)} {{\sigma ^2}}}} \right.\kern-\nulldelimiterspace} {{\sigma_r ^2}}}$ and ${\bf{\tilde Q}}_{{\bf{m}}\left[ n \right]}^{ - 1}$ only depends on the observation parameters ${\hat \theta_{\rm{e}}} \left[ n \right]$ and ${\hat \phi_{\rm{e}}} \left[ n \right]$. By introducing auxiliary variables ${v _i} \ge 0$, constraint \eqref{p1a} can be rewritten as
\begin{subequations}
	\label{ganzhi1}
	\begin{equation}
		\left[ {\begin{array}{*{20}{c}}
				{{{\bf{M}}^{ - 1}}\left[ n \right]}&{{{\bf{e}}_i}}\\
				{{\bf{e}}_i^{\rm{T}}}&{{v_i}}
		\end{array}} \right]\succeq0,i \in \left\{ {1, \ldots ,6} \right\},
	\end{equation}
	\begin{equation}
		\sum\limits_{i=1}^{6} v_i \le \Gamma_{\max},
	\end{equation}
	\begin{equation}
		v_i \ge 0,
	\end{equation}
\end{subequations}
where ${{\bf{e}}_i}$ is the $i$-th column of ${{\bf{I}}_6}$. For the non-convex constraint \eqref{qietingyueshu}, since $\left( {1 - {2^{{A_k}}}} \right)$ is negative, the majorization-minimization (MM) method is adopted to handle ${\left| {{{\bf{H}}_{\rm{e}}}\left[ n \right]{\bf{z}}\left[ n \right]} \right|^2}$ \cite{23}, which yields
\begin{align}
	\label{tag2}
	&\left| \mathbf{H}_{\mathrm{e}}[n]\mathbf{w}_k[n] \right|^2 \!+\! \left(1\! -\! 2^{A_k}\!\right) \Bigl(\!2\operatorname{Re}\bigl\{ \mathbf{z}_0^{\mathrm{H}} \mathbf{H}_{\mathrm{e}}[n] \mathbf{H}_{\mathrm{e}}^{\mathrm{H}}[n] (\mathbf{z}[n] - \mathbf{z}_0) \bigr\} \nonumber \\
	&+  \left| \mathbf{H}_{\mathrm{e}}[n]\mathbf{z}_0 \right|^2\Bigr) \le \left(2^{A_k}\! -\! 1\right) \sigma_{\mathrm{e}}^2,
\end{align}
where ${{\bf{z}}_0}$ denotes the initial point of ${\bf{z}}\left[ n \right]$. Therefore, problem \text{P3} can be transformed into
\begin{align}
	\text{P4:}~\operatorname*{max}_{\mathbf{w}[n],\,\mathbf{z}[n]} \ & F, \nonumber\\
	\text{s.t.}\quad & \eqref{ganzhi1}, \eqref{tag2},\eqref{p2c}. \nonumber
\end{align}
Problem \text{P4} is a convex optimization problem and can be efficiently solved using standard convex optimization solvers, such as CVX \cite{9982476}.

\subsubsection{Passive Beamforming Design for Secure Communication Signals}

For subproblem 2, given ${\bf w}[n]$, ${\bf z}[n]$, ${\bf \Psi}[n]$, and $A_k$, the RIS reflection coefficients for adjusting the secure communication signal are optimized. Thus, problem $\text{P2}$ can be reformulated with respect to ${\bf f}[n]$ as
\begin{subequations}
	\begin{align}
		\text{P5:}~\operatorname*{max}_{{\bf{f}}\left[ n \right]}\quad& F, \nonumber \\
		\text{s.t.}\quad
		&\eqref{p1a}, \eqref{qietingyueshu}, \eqref{p2d}. \nonumber 
	\end{align}
\end{subequations}
Due to the coupled quadratic form in $\mathbf f[n]$ and the unit-modulus constraint, the resulting subproblem is difficult to handle directly. We therefore apply the MM algorithm and upper bound the quadratic term $\mathbf f^{\rm{H}}[n]\mathbf D\mathbf f[n]$ by a tight first order surrogate at $\mathbf f_0$ \cite{32}, which gives
\setcounter{equation}{49}
\begin{align}
	\label{songchi1}
	&{{\bf{f}}^{\rm{H}}}\left[ n \right]{\bf{Df}}\left[ n \right] \le 2{\mathop{\rm Re}\nolimits} \left\{ {{{\left( {{{\bf{f}}_0}} \right)}^{\rm{H}}}\left( {{\bf{D}} - \left\| {\bf{D}} \right\|_F^2{{\bf{I}}_M}} \right){\bf{f}}\left[ n \right]} \right\} \nonumber \\
	&+ M\left\| {\bf{D}} \right\|_F^2 - {\left( {{{\bf{f}}_0}} \right)^{\rm{H}}}\left( {{\bf{D}} - \left\| {\bf{D}} \right\|_F^2{{\bf{I}}_M}} \right){{\bf{f}}_0},
\end{align}
where $\mathbf f_0$ is the feasible point from the previous iteration. For the non-convex constraint \eqref{p1a}, we first rewrite the SNR with respect to ${\bf{f}}\left[ n \right]$ as
\begin{align}
	\frac{{{{\left| {{\bf{H}}\left[ n \right]{\bf{x}}\left[ n \right]} \right|}^2}}}{{{\sigma_r ^2}}} = \frac{{{{\bf{f}}^{\rm{H}}}\left[ n \right]{\bf{\Xi f}}\left[ n \right]}}{{{\sigma_r ^2}}}, 
\end{align}
where ${\bf{\Xi }}$ is given by \eqref{DL SINR}. To obtain a tractable surrogate of \eqref{p1a}, we apply the MM algorithm as
\begin{figure*}[ht]
	\normalsize
	\centering
	\begin{align}\label{DL SINR}
		\begin{aligned}
		\boldsymbol{\Xi}
		=
		\left(
		\mathbf h_{\rm re}^{\rm H}[n]
		\boldsymbol{\Psi}^{\rm H}[n]
		\mathbf H_{\rm RB}^{\rm H}
		\mathbf H_{\rm RB}
		\boldsymbol{\Psi}[n]
		\mathbf h_{\rm re}[n]
		\right)
		\times
		\Bigg[
		\mathbf h_{\rm re}[n]\mathbf h_{\rm re}^{\rm H}[n]
		\odot
		\Bigg(
		\mathbf H_{\rm BR}
		\left(
		\sum_{k=1}^{K}
		\mathbf w_k[n]\mathbf w_k^{\rm H}[n]
		+
		\mathbf z[n]\mathbf z^{\rm H}[n]
		\right)
		\mathbf H_{\rm BR}^{\rm H}
		\Bigg)^{\rm T}
		\Bigg].
		\end{aligned}
	\end{align}
	\hrulefill
\end{figure*}
\begin{align}
	&{{\bf{f}}^{\rm{H}}}\left[ n \right]{\bf{\Xi f}}\left[ n \right] \ge {\mathop{\rm Re}\nolimits} \left\{ {2{{\left( {{{\bf{f}}_0}} \right)}^{\rm{H}}}{{\bf{\Xi }}^{\rm{H}}}{\bf{f}}\left[ n \right]} \right\} - {\mathop{\rm Re}\nolimits} \left\{ {2{{\left( {{{\bf{f}}_0}} \right)}^{\rm{H}}}{{\bf{\Xi }}^{\rm{H}}}{{\bf{f}}_0}} \right\}\nonumber \\
	&+{\left( {{{\bf{f}}_0}} \right)^{\rm{H}}}{\bf{\Xi }}{{\bf{f}}_0}.
\end{align}
For the non-convex constraint \eqref{qietingyueshu}, we rewrite it with respect to ${\bf{f}}\left[ n \right]$ as
\begin{equation}
	\label{tag3}
\!\!	{{\bf{f}}^{\rm{H}}}\left[ n \right]{{\bf{D}}_3}{\bf{f}}\left[ n \right] \!+\! \left( {1\! -\! {2^{{A_k}}}} \right){{\bf{f}}^{\rm{H}}}\left[ n \right]{{\bf{D}}_4}{\bf{f}}\left[ n \right] \le \left( {{2^{{A_k}}}\! -\! 1} \right)\sigma _{\rm{e}}^2,
\end{equation}
where $\mathbf D_3
=
\left(
\mathbf h_{\rm re}^{\rm H}[n]
\operatorname{diag}
\left(
\mathbf H_{\rm BR}\mathbf w_k[n]
\right)
\right)^{\rm H}
(
\mathbf h_{\rm re}^{\rm H}[n]
\operatorname{diag}
(
\mathbf H_{\rm BR} 
\\ \mathbf w_k[n]
)
)$ and $\mathbf D_4
=\left(
\mathbf h_{\rm re}^{\rm H}[n]
\operatorname{diag}
\left(
\mathbf H_{\rm BR}\mathbf z[n]
\right)
\right)^{\rm H}
(
\mathbf h_{\rm re}^{\rm H}[n]
\operatorname{diag}\\
(
\mathbf H_{\rm BR}\mathbf z[n]
)
)$. The two quadratic terms in \eqref{tag3} are approximated by the MM method as
\begin{align}
	&{{\bf{f}}^{\rm{H}}}\left[ n \right]{{\bf{D}}_3}{{\bf{f}}}\left[ n \right] \le 2{\mathop{\rm Re}\nolimits} \left\{ {{{\left( {{{\bf{f}}_0}} \right)}^{\rm{H}}}\left( {{{\bf{D}}_3} - \left\| {{{\bf{D}}_3}} \right\|_F^2{{\bf{I}}_M}} \right){\bf{f}}\left[ n \right]} \right\} \nonumber \\
	&+ M\left\| {{{\bf{D}}_3}} \right\|_F^2 - {\left( {{{\bf{f}}_0}} \right)^{\rm{H}}}\left( {{{\bf{D}}_3} - \left\| {{{\bf{D}}_3}} \right\|_F^2{{\bf{I}}_M}} \right){{\bf{f}}_0},
\end{align}
\begin{align}
	&{{\bf{f}}^{\rm{H}}}\left[ n \right]{{\bf{D}}_4}{\bf{f}}\left[ n \right] \ge {\mathop{\rm Re}\nolimits} \left\{ {2{{\left( {{{\bf{f}}_0}} \right)}^{\rm{H}}}{{\bf{D}}_4}^{\rm{H}}{\bf{f}}\left[ n \right]} \right\} + {\left( {{{\bf{f}}_0}} \right)^{\rm{H}}}{{\bf{D}}_4}{{\bf{f}}_0} \nonumber \\
	&- {\mathop{\rm Re}\nolimits} \left\{ {2{{\left( {{{\bf{f}}_0}} \right)}^{\rm{H}}}{{\bf{D}}_4}^{\rm{H}}{{\bf{f}}_0}} \right\}.
\end{align}

Then, the unit-modulus constraint of ${\bf{f}}\left[ n \right]$ is handled by the ADMM algorithm. Specifically, we introduce an auxiliary variable ${\boldsymbol{\varphi }} \buildrel \Delta \over = {\left[ {{\varphi _1},{\varphi _2}, \ldots ,{\varphi _M}} \right]^{\rm{T}}}$ and transform \eqref{p2d} into
\begin{subequations}
	\label{xiaxing}
	\begin{equation}
		\left| {{{\left[ {{\bf{f}}\left[ n \right]} \right]}_m}} \right| \le 1,
	\end{equation}
	\begin{equation}
		{\bf{f}}\left[ n \right] = {\boldsymbol{\varphi }}.
	\end{equation}
\end{subequations}
Based on the ADMM algorithm, the optimization problem is transformed into the following augmented Lagrangian form:
\begin{align}
	\label{dierge}
	\text{P6:}~\operatorname*{max}_{\mathbf{f}[n]}\quad
	& F - \frac{1}{2\zeta}\left\| \mathbf{f}[n] - \boldsymbol{\varphi} + \zeta \boldsymbol{\chi} \right\|^2, \nonumber\\
	\text{s.t.}\quad
	& \eqref{ganzhi1},\ \eqref{xiaxing}, \nonumber\\
	& M\left\| \mathbf{D}_3 \right\|_F^2
	+ 2\operatorname{Re}\bigl\{ \mathbf{f}_0^{\mathrm{H}} (\mathbf{D}_3 - \left\| \mathbf{D}_3 \right\|_F^2 \mathbf{I}_M) \mathbf{f}[n] \bigr\} \nonumber\\
	& - \mathbf{f}_0^{\mathrm{H}} (\mathbf{D}_3 - \left\| \mathbf{D}_3 \right\|_F^2 \mathbf{I}_M) \mathbf{f}_0 \nonumber\\
	& + (1 - 2^{A_k}) \Bigl( \mathbf{f}_0^{\mathrm{H}} \mathbf{D}_4 \mathbf{f}_0
	+ \operatorname{Re}\{ 2 \mathbf{f}_0^{\mathrm{H}} \mathbf{D}_4^{\mathrm{H}} \mathbf{f}[n] \} \nonumber\\
	& - \operatorname{Re}\{ 2 \mathbf{f}_0^{\mathrm{H}} \mathbf{D}_4^{\mathrm{H}} \mathbf{f}_0 \} \Bigr)
	\le (2^{A_k} - 1)\sigma_{\mathrm{e}}^2, 
\end{align}
where ${\boldsymbol{\chi}}\in\mathbb{C}^{M}$ denotes the dual variable, and $\zeta  > 0$ is a preset penalty parameter. For fixed ${\boldsymbol{\varphi}}$ and ${\boldsymbol{\chi}}$, the update of ${\bf{f}}\left[ n \right]$ is convex. For fixed ${\bf{f}}\left[ n \right]$ and ${\boldsymbol{\chi }}$, the optimal auxiliary variable ${\boldsymbol{\varphi}}$ is obtained by phase alignment as
\begin{equation}
	{{\boldsymbol{\varphi }}^ \star } = {e^{j\angle \left( {{\bf{f}}\left[ n \right] + \zeta {\boldsymbol{\chi }}} \right)}}.
\end{equation}
After obtaining ${\bf{f}}\left[ n \right]$ and ${\boldsymbol{\varphi }}$, the dual variable ${\boldsymbol{\chi }}$ is updated by
\begin{equation}
	{\boldsymbol{\chi }}: = {\boldsymbol{\chi }} + {{\left( {{\bf{f}}\left[ n \right] - {\boldsymbol{\varphi }}} \right)} \mathord{\left/
			{\vphantom {{\left( {{\bf{f}}\left[ n \right] - {\bf{\varphi }}} \right)} \zeta }} \right.
			\kern-\nulldelimiterspace} \zeta }.
\end{equation}

\subsubsection{Passive Beamforming Design for Echo Sensing Signals}

For subproblem 3, given ${\bf{w}}\left[ n \right]$, ${\bf{z}}\left[ n \right]$, and ${\bf{f}}\left[ n \right]$, the RIS coefficients for the echo sensing signal are obtained by satisfying the tracking accuracy constraint and the unit-modulus constraint. The corresponding feasibility problem is given by
\begin{subequations}
	\begin{align}
		\text{P7:}\quad~&\text{find}~{\bf{\Psi }}\left[ n \right],\nonumber \\
		\text{s.t.}\quad
		&\eqref{p1a}, \eqref{p1e}. \nonumber
	\end{align}
\end{subequations}
Let ${\bf{p}}\left[ n \right] = {\rm{diag}}\left( {{\bf{\Psi }}\left[ n \right]} \right)$. For the non-convex constraint \eqref{p1a}, the SNR can be rewritten with respect to ${\bf{p}}\left[ n \right]$ as
\setcounter{equation}{60}
\begin{equation}
	\frac{{{{\left| {{\bf{H}}\left[ n \right]{\bf{x}}\left[ n \right]} \right|}^2}}}{{{\sigma_r ^2}}} = \frac{{{{\bf{p}}^{\rm{H}}}\left[ n \right]{{\bf{\Xi }}_1}{\bf{p}}\left[ n \right]}}{{{\sigma_r ^2}}},
\end{equation}
where ${{\bf{\Xi }}_1}$ is written as \eqref{UL SINR}. Similarly, the MM algorithm is adopted to handle it as
\begin{figure*}[!b]
	\normalsize
	\hrulefill 
	\centering
	\begin{align}\label{UL SINR}
		\begin{aligned}
			\boldsymbol{\Xi}_1
			=
			\Bigg(
			\mathbf h_{\rm re}^{\rm H}[n]
			\boldsymbol{\Theta}[n]
			\mathbf H_{\rm BR}
			\left(
			\sum_{k=1}^{K}
			\mathbf w_k[n]\mathbf w_k^{\rm H}[n]
			+
			\mathbf z[n]\mathbf z^{\rm H}[n]
			\right)
			\mathbf H_{\rm BR}^{\rm H}
			\boldsymbol{\Theta}^{\rm H}[n]
			\mathbf h_{\rm re}[n]
			\Bigg)
			\times
			\left[
			\mathbf H_{\rm RB}^{\rm H}
			\mathbf H_{\rm RB}
			\odot
			\left(
			\mathbf h_{\rm re}[n]\mathbf h_{\rm re}^{\rm H}[n]
			\right)^{\rm T}
			\right].
		\end{aligned}
	\end{align}
\end{figure*}

\begin{equation}
\!\!	\mathbf p^{\rm{H}}[n]\boldsymbol{\Xi}_{1}\mathbf p[n]
	\geq
	2\operatorname{Re}\left\{
	\mathbf p_{0}^{\rm{H}}[n]\boldsymbol{\Xi}_{1}\mathbf p[n]
	\right\}
	-
	\mathbf p_{0}^{\rm{H}}[n]\boldsymbol{\Xi}_{1}\mathbf p_{0}[n].
\end{equation}
For the unit-modulus constraint of ${\bf{p}}\left[ n \right]$, we relax \eqref{p1e} as
\begin{equation}
	\label{tag4}
	\left| {{{\left[ {{\bf{p}}\left[ n \right]} \right]}_m}} \right| \le 1.
\end{equation}
After solving ${\bf{p}}\left[ n \right]$, a normalization operation is performed to recover the unit-modulus structure. Therefore, problem \text{P7} can be transformed into
\begin{subequations}
	\label{disange}
	\begin{align}
		\text{P8:}\quad~&\text{find}~{\bf{p}}\left[ n \right],\nonumber \\
		\text{s.t.}\quad
		&\eqref{ganzhi1},\eqref{tag4}. \nonumber
	\end{align}
\end{subequations}
Problem \text{P8} is a convex feasibility problem and can be efficiently solved by standard convex optimization solvers such as CVX. After the above block updates, the eavesdropping rate upper bound is directly updated according to the active condition of \eqref{p1b}.

\subsection{Convergence and Computational Complexity Analysis}

\begin{algorithm}[!t]
	\caption{Joint Beamforming Optimization and Dynamic Tracking Algorithm}
	\label{suanfa1}
	\begin{algorithmic}[1]
		\Require Iteration number $l=1$, time slot number $n=1$ and convergence threshold $\Gamma$.
		\Repeat
		\State Initialize feasible points $\mathbf{w}_0$, $\mathbf{z}_0$, $\mathbf{f}_0$, $\mathbf{p}_0$, $\{A_k^0\}$ and
		\StateContm  penalty factor $\zeta$.
		\Repeat
		\State Calculate auxiliary variables ${\mathbf{r}}^{l}$ and ${\mathbf{c}}^{l}$ via \eqref{gengxin1} and
		\StateCont \eqref{gengxin2}.
		\State Given ${\mathbf{f}}^{l-1}[n]$, ${\mathbf{p}}^{l-1}[n]$, $\{A_k^{l-1}\}$, ${\mathbf{r}}^{l}$ and ${\mathbf{c}}^{l}$, update 
		\StateCont ${\mathbf{w}}^{l}[n]$ and ${\mathbf{z}}^{l}[n]$ by solving problem $\text{P4}$.
		\State Given ${\mathbf{w}}^{l}[n]$, ${\mathbf{z}}^{l}[n]$, ${\mathbf{p}}^{l-1}[n]$, $\{A_k^{l-1}\}$, ${\mathbf{r}}^{l}$ and ${\mathbf{c}}^{l}$,  
		\StateCont update ${\mathbf{f}}^{l}[n]$ by solving problem $\text{P6}$.
		\State Given ${\mathbf{w}}^{l}[n]$, ${\mathbf{z}}^{l}[n]$, and ${\mathbf{f}}^{l}[n]$, update 
		\StateCont ${\mathbf{p}}^{l}[n]$ by solving problem $\text{P8}$.
		\State Update $\{A_k^{l}\}$.
		\State Update $l = l + 1$.
		\Until{The fractional diminution of objective function}
		\StateContm  value falls below threshold $\Gamma$. The obtained joint 
		\StateContm beamforming design for the $n$-th time slot $\mathbf{w}^\star[n]$,
		\StateContm  $\mathbf{z}^\star[n]$, $\mathbf{f}^\star[n]$, and $\mathbf{p}^\star[n]$ can be saved. Reset $l=1$.
		\State Update $n = n + 1$.
		\Until{Iteration number $n = N$}.
		\Ensure All time slots joint beamforming designs $\{\mathbf{w}^\star[n]\}$,
		\StateContn  $\{\mathbf{z}^\star[n]\}$, $\{\mathbf{f}^\star[n]\}$ and $\{\mathbf{p}^\star[n]\}$.
	\end{algorithmic}
\end{algorithm}

The joint optimization algorithm of beamforming and dynamic tracking is summarized in \textbf{Algorithm} \ref{suanfa1}. In the following, we analyze its convergence and computational complexity.
\subsubsection{Computational Complexity Analysis}

The computational complexity of \textbf{Algorithm}~\ref{suanfa1}
mainly arises from solving $\mathrm{P4}$, $\mathrm{P6}$, and
$\mathrm{P8}$ in each outer iteration and each time slot.
Specifically, $\mathrm{P4}$ and $\mathrm{P8}$ are solved by the
primal dual interior point method, whereas $\mathrm{P6}$ is handled
by ADMM, in which the update of $\mathbf f[n]$ requires solving a
convex subproblem. For a generic conic program with decision dimension
$d$ and barrier parameter $\nu$, the computational complexity is
approximately
$\mathcal{O}\!\left(\sqrt{\nu}\log(1/\epsilon)d^{3}\right)$.
Since the decision dimensions of $\mathrm{P4}$, the update of
$\mathbf f[n]$ in $\mathrm{P6}$, and $\mathrm{P8}$ scale as
$\mathcal{O}\!\left(N_t(K+1)\right)$, $\mathcal{O}(M)$, and
$\mathcal{O}(M)$, respectively, their complexities are
$\mathcal{O}\!\left(\sqrt{\nu_{4}}\log(1/\epsilon)
N_t^{3}(K+1)^{3}\right)$,
$\mathcal{O}\!\left(I_{\rm ADMM}\sqrt{\nu_{6}}
\log(1/\epsilon)M^{3}\right)$, and
$\mathcal{O}\!\left(\sqrt{\nu_{8}}\log(1/\epsilon)M^{3}\right)$,
where $I_{\rm ADMM}$ denotes the average number of ADMM iterations.
Define
$\eta_s\triangleq I_{\rm ADMM}\sqrt{\nu_6}+\sqrt{\nu_8}$.
The closed form updates of the auxiliary and dual variables, as well
as the EKF operations with a fixed state dimension, have lower
computational orders and are omitted. Therefore, letting
$I_{\rm out}$ denote the average number of outer iterations per time
slot, the overall computational complexity is
\[
	\mathcal{O}\!\left(
	NI_{\rm out}\log(1/\epsilon)
	\left[
	\sqrt{\nu_{4}}N_t^{3}(K+1)^{3}
	+\eta_sM^{3}
	\right]
	\right).\]

\subsubsection{Convergence Analysis}
The convergence of proposed \textbf{Algorithm} \ref{suanfa1} can be shown as follows. We define the objective function as ${\mathcal{R}} \left( {{{\bf{w}}^l}\left[ n \right],{{\bf{z}}^l}\left[ n \right],{{\bf{f}}^l}\left[ n \right],{{\bf{p}}^l}\left[ n \right]} \right)$, where ${{\bf{w}}^l}\left[ n \right]$, ${{\bf{z}}^l}\left[ n \right]$, ${{\bf{f}}^l}\left[ n \right]$, ${{\bf{p}}^l}\left[ n \right]$ denotes the solution to problems \text{P4}, \text{P6}, and \text{P8} at the $l$-th iteration. Based on the aforementioned derivation process, we can obtain
\setcounter{equation}{64}
\begin{align}
	&{\mathcal{R}} \left( {{{\bf{w}}^l}\left[ n \right],{{\bf{z}}^l}\left[ n \right],{{\bf{f}}^l}\left[ n \right],{{\bf{p}}^l}\left[ n \right]} \right) \nonumber \\
	&\mathop  \le \limits^a {\mathcal{R}} \left( {{{\bf{w}}^{l + 1}}\left[ n \right],{{\bf{z}}^{l + 1}}\left[ n \right],{{\bf{f}}^l}\left[ n \right],{{\bf{p}}^l}\left[ n \right]} \right) \nonumber \\
	&\mathop  \le \limits^b {\mathcal{R}} \left( {{{\bf{w}}^{l + 1}}\left[ n \right],{{\bf{z}}^{l + 1}}\left[ n \right],{{\bf{f}}^{l + 1}}\left[ n \right],{{\bf{p}}^l}\left[ n \right]} \right) \nonumber \\
	&\mathop  \le \limits^c {\mathcal{R}} \left( {{{\bf{w}}^{l + 1}}\left[ n \right],{{\bf{z}}^{l + 1}}\left[ n \right],{{\bf{f}}^{l + 1}}\left[ n \right],{{\bf{p}}^{l + 1}}\left[ n \right]} \right),
\end{align}
which shows the objective value is non-decreasing across alternating-optimization iterations. Specifically, at iteration $l$ for time slot $n$, \textbf{Algorithm} \ref{suanfa1} performs block-wise sequential updates. With \({\bf f}^{l}[n]\), \({\bf p}^{l}[n]\), and \(\{A_k^{l}\}\) fixed, step \text{5} solves problem \text{P4} to update ISAC BS beamforming and AN (\(\bf w^{l+1}[n], \bf z^{l+1}[n]\)), which leads to inequality $a$. Then, given \({\bf w}^{l+1}[n],{\bf z}^{l+1}[n],{\bf p}^{l}[n]\), and \(\{A_k^{l}\}\), step \text{6} optimizes DL RIS reflection coefficients via problem \text{P6} to obtain \({\bf f}^{l+1}[n]\), thus establishing inequality $b$. Likewise, step \text{7} updates UL RIS coefficients by solving feasibility problem \text{P8} for \({\bf p}^{l+1}[n]\), which results in inequality~$c$, satisfying UL sensing constraints. The algorithm thus generates a monotonically non-decreasing objective sequence. Moreover, since ISAC BS transmit power is bounded and the involved noise powers are finite, the achievable total secrecy rate is upper bounded.

\section{Numerical Results} \label{sibufen}
\begin{table}[!t]
	\centering
	\refstepcounter{table}
	\label{tab1}
	{\footnotesize TABLE~\thetable\par}
	\vspace{1mm}
	{\footnotesize\scshape Simulation Parameters\par}
	\vspace{1.5mm}
	\renewcommand{\arraystretch}{1.18}
	\setlength{\tabcolsep}{3pt}
	\footnotesize
	\begin{tabular}{|>{\centering\arraybackslash}m{0.58\columnwidth}|
			>{\centering\arraybackslash}m{0.32\columnwidth}|}
		\hline
		\textbf{Parameters} & \textbf{Value} \\
		\hline

		Number of BS transmit antennas & $N_t=16$ \\
		\hline
		
		Number of BS receive antennas & $N_r=12$ \\
		\hline
		
		Number of users & $K=3$ \\
		\hline
		
		Number of RIS elements per region & $M=36$ \\
		\hline
		
		Number of time slots & $N=14$ \\
		\hline
		
		Maximum BS transmit power & $P_{\max}=35$ dBm \\
		\hline
		
		Carrier frequency & $f_c=30$ GHz \\
		\hline
		
		Antenna/RIS element spacing & $\lambda/2$ \\
		\hline
		
		Locations of the BS and RIS 
		& \makecell[c]{
			\rule{0pt}{2.8ex}$\mathbf q_{\rm bs}=[0,0,10]^T$ m \\
			$\mathbf q_{\rm r}=[20,20,13]^T$ m
			\rule[-1.2ex]{0pt}{1.2ex}
		} \\
		\hline
		
		Locations of users
		& \makecell[c]{\rule{0pt}{2.8ex}
			$\mathbf q_1=[37,35,0]^T$ m \\
			$\mathbf q_2=[33,37,0]^T$ m \\
			$\mathbf q_3=[32,31,0]^T$ m
			\rule[-1.2ex]{0pt}{1.2ex}
		} \\
		\hline
		
		Initial eavesdropper position 
		\rule{0pt}{2.8ex}
		& $[31.5,33.0,1.50]^T$ m
		\rule[-1.2ex]{0pt}{1.2ex} \\
		\hline
		
		\makecell[c]{Path-loss exponents of the BS-RIS,\\
			RIS-user, and RIS-eavesdropper links}
		& $2.0,\ 2.2,\ 2.35$ \\
		\hline
		
		\makecell[c]{Rician factors of the BS-RIS,\\
			and RIS-user links}
		& $0$ dB, $8$ dB \\
		\hline
		
		 Reference path loss 
		 & $C_0=-20$ dB \\
		 \hline
		
		 Reference distance 
		 & $D_0=1$ m \\
		 \hline
		
		\makecell[c]{Noise powers at the user,\\
			eavesdropper, and BS receiver}
		& \makecell[c]{
			$-90$ dBm, $-85$ dBm, \\
			$-100$ dBm
		} \\
		\hline
		
		Tracking accuracy threshold 
		& $\Gamma_{\max}=0.075$ \\
		\hline

	\end{tabular}
\end{table}

In this section, numerical simulations are conducted to evaluate the 
performance of the proposed algorithm. The BS is equipped with $N_t=16$ transmit antennas and $N_r=12$ receive antennas, and serves $K=3$ users with the aid
of the RIS. Each RIS region contains $M=36$ elements, 
and the service period is divided into $N=14$ time slots. The maximum BS 
transmit power and carrier frequency are set to $P_{\max}=35$ dBm and 
$f_c=30$ GHz, respectively, while the antenna and RIS element spacings are 
both $\lambda/2$. The BS and RIS are located at 
$\mathbf q_{\rm bs}=[0,0,10]^T$ m and 
$\mathbf q_{\rm r}=[20,20,13]^T$ m, respectively. The user locations are 
$\mathbf q_1=[37,35,0]^T$ m, $\mathbf q_2=[33,37,0]^T$ m, and 
$\mathbf q_3=[32,31,0]^T$ m, while the initial eavesdropper position is 
$[31.5,33.0,1.50]^T$ m. The large-scale attenuation of each RIS-related 
link follows the distance-dependent power-law model
$L(d)=C_0(d/D_0)^{-\alpha}$, where $C_0=-20$ dB is the reference channel 
power gain at $D_0=1$ m and $\alpha$ denotes the corresponding path-loss 
exponent. The path-loss exponents of the BS-RIS, RIS-user, and RIS-eavesdropper links are set to 2.0, 2.2, and 2.35, respectively, and the corresponding Rician factors are 0 dB, 8 dB, and 12 dB. 
The noise powers at the users, eavesdropper, and BS receiver are 
-90 dBm, -85 dBm, and -100 dBm, respectively, and the tracking 
accuracy threshold is $\Gamma_{\max}=0.075$. The main simulation parameters 
are summarized in Table~\ref{tab1}. Based on these settings, we evaluate the 
effects of the RIS size, BS transmit power, number of BS antennas, and 
tracking accuracy requirement on the achievable secrecy rate.

\begin{figure}
	\centering
	\includegraphics[width=\columnwidth]{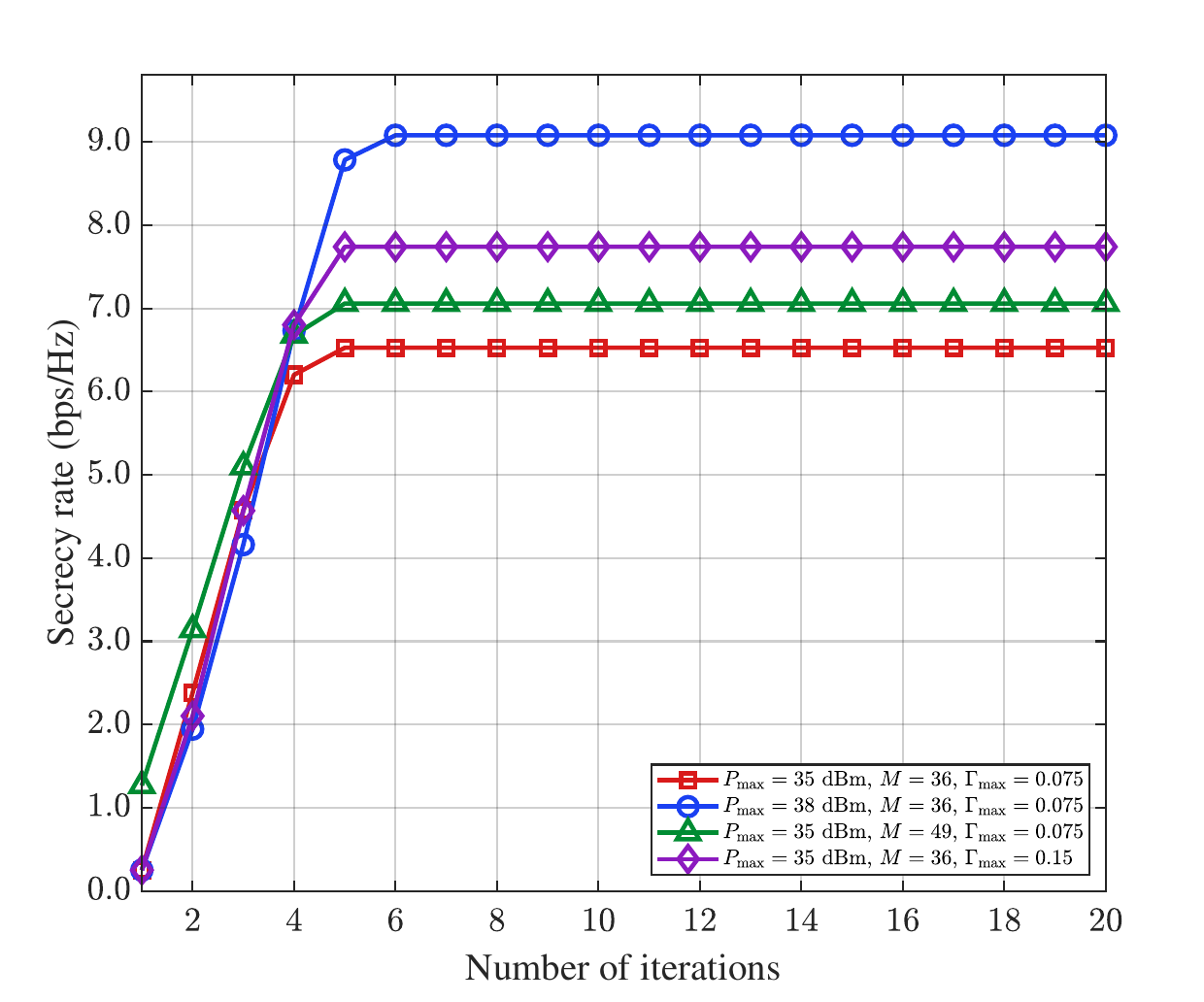}
	\caption{Convergence behavior of the proposed algorithm.}
	\label{convergence}
\end{figure}

We first characterize the convergence performance of the proposed optimization algorithm in Fig.~\ref{convergence}, which plots the evolution of the system total secrecy rate versus the number of iterations under different BS transmit power budgets, different RIS sizes, and different sensing accuracy requirements. It is observed that the total secrecy rate rises rapidly at the initial iterations and converges within a small number of iterations for all considered parameter configurations, which verifies the effectiveness and favorable convergence behavior of our proposed algorithm.

\begin{figure}
	\centering
	\includegraphics[width=\columnwidth]{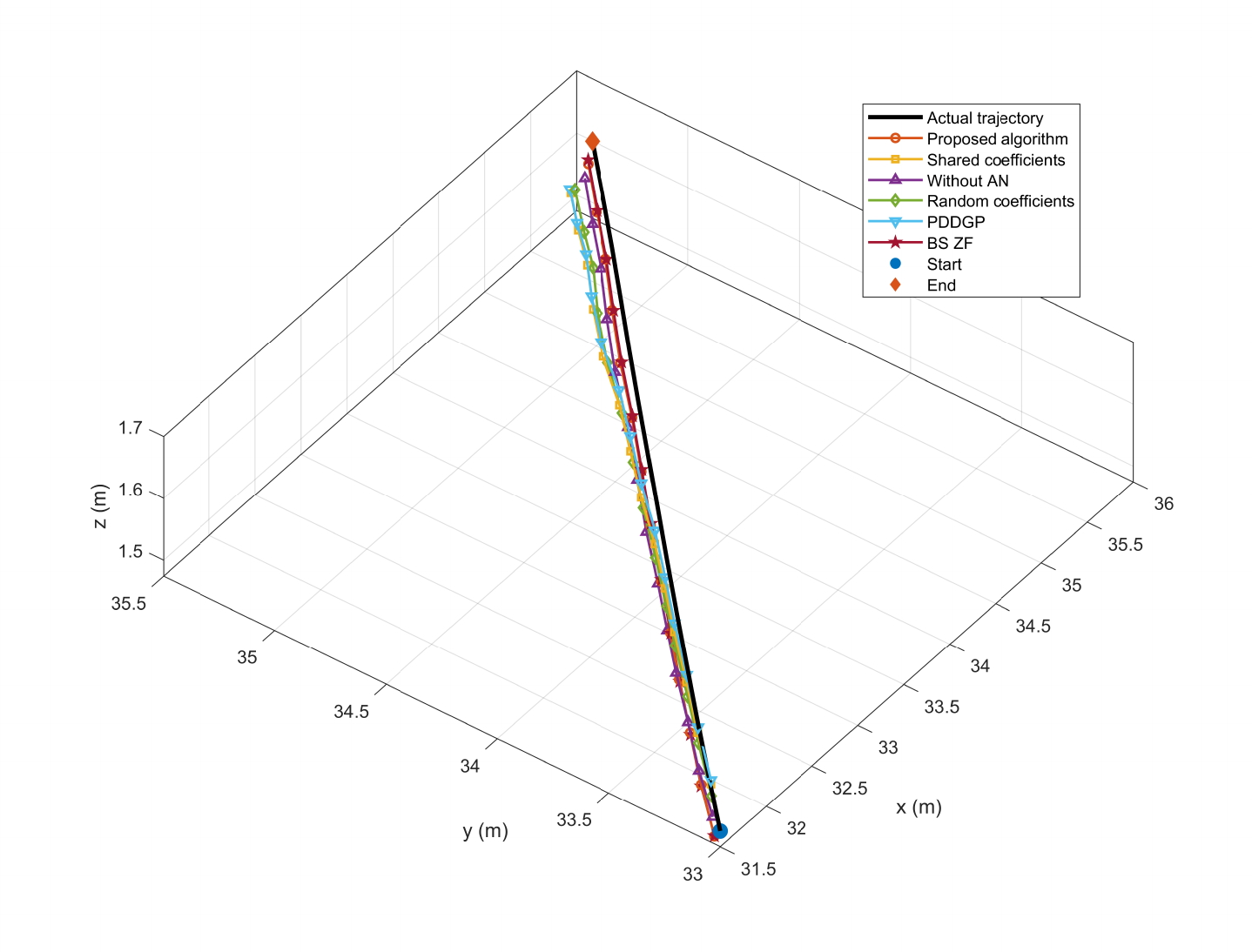}
	\caption{Comparison of tracking trajectories for the eavesdropper.}
	\label{trajectory}
\end{figure}

We then compare the simulated eavesdropper trajectory tracking performance of the proposed algorithm against multiple benchmark schemes, which are elaborated as follows. (1) Shared coefficients: This benchmark adopts identical RIS reflection coefficients for secure communication and echo sensing, i.e., $\mathbf p[n]=\mathbf f[n]$. (2) Without AN: AN is eliminated from the BS transmit signal in this scheme, which highlights the performance contribution of AN in suppressing wiretap channels. (3) Random coefficients: The RIS reflection coefficients dedicated to echo sensing are randomly generated with unit modulus. (4) Penalty Dual Decomposition Gradient Projection (PDDGP) \cite{9790808}: The PDDGP framework is adapted to our considered system model. Specifically, the BS beamforming and AN vectors are first updated under identical constraints. Afterwards, the RIS reflection coefficients for secure transmission and echo sensing are separately optimized via low-complexity projected gradient iterations, followed by projecting each RIS phase vector onto the unit-modulus feasible region. This benchmark is utilized to compare the performance gap between our proposed algorithm and low-complexity RIS phase optimization strategies. (5) BS Zero-Forcing (ZF): BS transmit beamforming is constructed following the conventional ZF criterion. As observed from Fig.~\ref{trajectory}, all comparative schemes can roughly track the mobility of the eavesdropper, while the trajectory yielded by our proposed algorithm achieves the minimum deviation from the ground-truth eavesdropper path.

\begin{figure}
	\centering
	\includegraphics[width=\columnwidth]{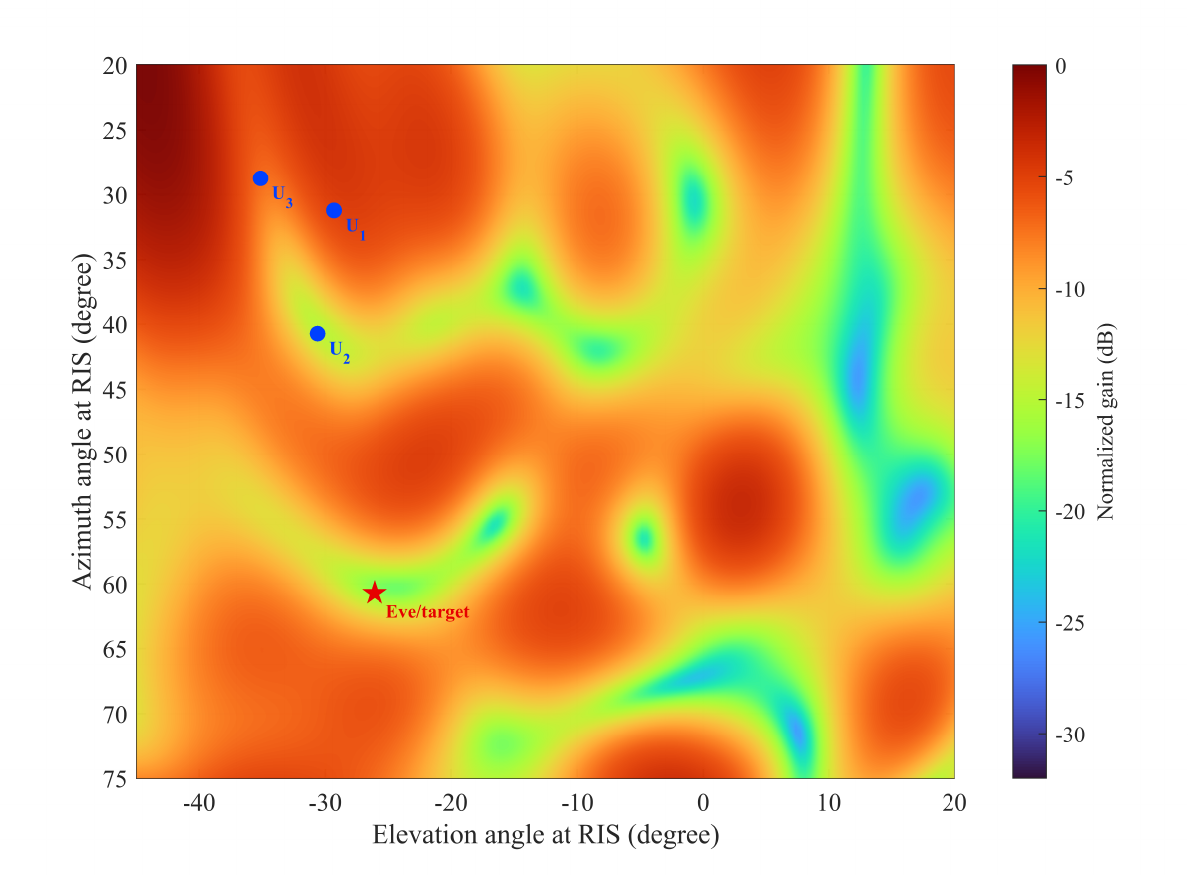}
	\caption{Normalized beam-gain distribution of the ISAC signal.}
	\label{ISAC signal}
\end{figure}

\begin{figure}
	\centering
	\includegraphics[width=\columnwidth]{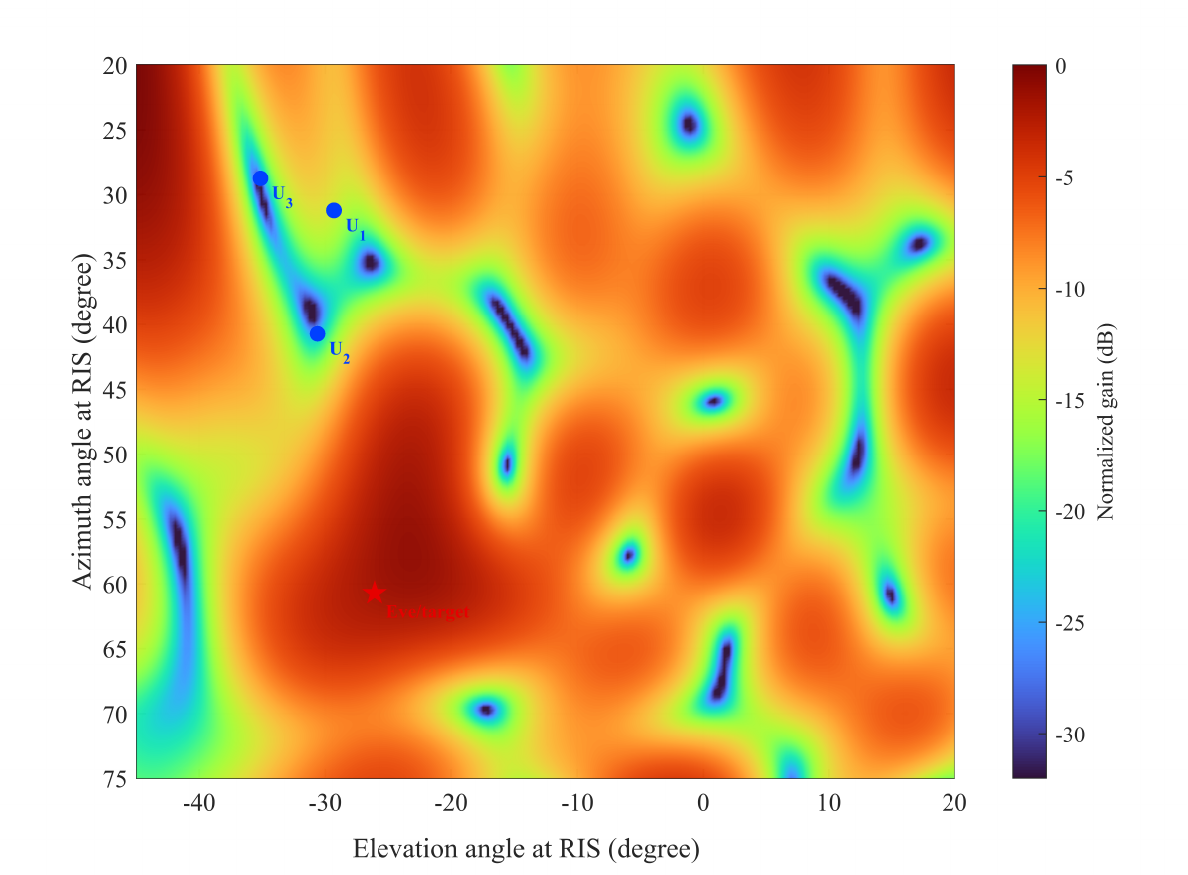}
	\caption{Normalized beam-gain distribution of the AN signal.}
	\label{AN}
\end{figure}

Figs.~\ref{ISAC signal} and~\ref{AN} depict the normalized beam patterns of the ISAC signal and AN within a representative time slot, respectively. It can be observed that the ISAC signal attains considerable radiation gains toward each user, which boosts the received power of desired information signals at user terminals. In contrast, the AN beam exhibits low-gain regions covering all user directions, revealing that the optimized AN avoids deteriorating the SINR of users. Meanwhile, the AN energy is concentrated around the eavesdropper’s angular range with prominent beam gains, which generates deliberate interference over potential wiretap regions. The two beam patterns collectively verify that the proposed algorithm achieves spatial decoupling between confidential information transmission and AN jamming. This further demonstrates that our developed framework is capable of elevating the system secrecy rate while complying with the tracking constraints imposed by the sensing task.

\begin{figure}
	\centering
	\includegraphics[width=\columnwidth]{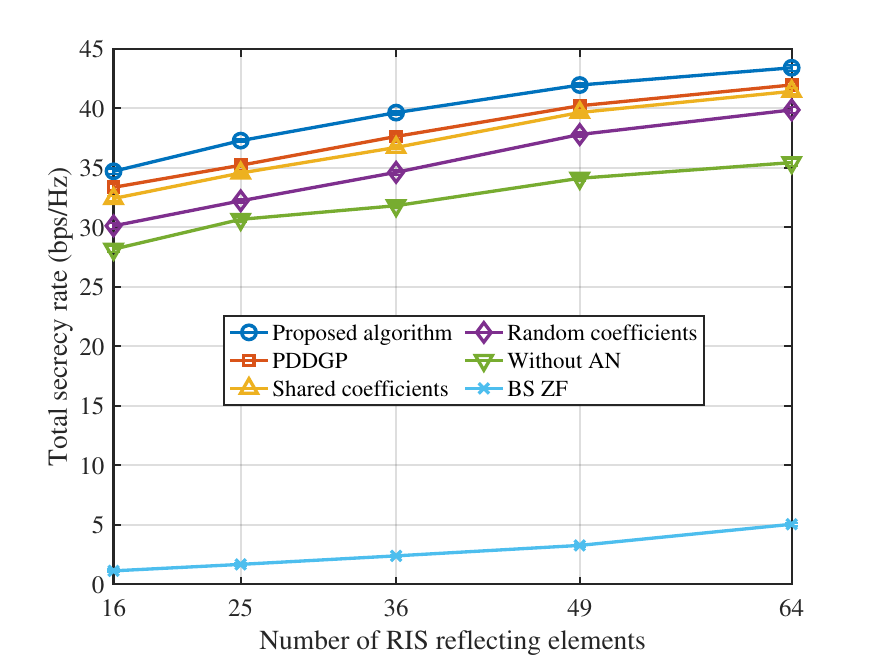}
	\caption{Total secrecy rate versus the number of RIS reflecting elements.}
	\label{RIS number}
\end{figure}

We next investigate the influence of the RIS size on the secrecy performance, as shown in Fig.~\ref{RIS number}. To justify the adopted RIS size, we conduct a LoS-based link-budget analysis for the cascaded sensing channel. To facilitate estimation, we adopt the initial position of the eavesdropper and ignore the distance variation between the eavesdropper and the RIS over different time slots. Under ideal coherent phase alignment, the maximum sensing SNR is given by
\begin{equation}
	\begin{aligned}
		S
		=\frac{P_{\max}N_tN_rM^4C_0^4}{\sigma_r^2}
		\left(\frac{\kappa}{1+\kappa}\right)^2
		\left(\frac{D_0^2}{d_{\mathrm{dr}}d_{\mathrm{re}}}\right)^4.
	\end{aligned}
	\label{eq:ris_sensing_snr}
\end{equation}
Accordingly, the minimum RIS size required to achieve a sensing SNR of $S_0$ is
\begin{equation}
	\begin{aligned}
		M_{\min}=
		\frac{d_{\mathrm{dr}}d_{\mathrm{re}}}{D_0^2}
		\left(\frac{S_0\sigma_r^2}
		{P_{\max}N_tN_rC_0^4}\right)^{\frac14}
		\sqrt{\frac{1+\kappa}{\kappa}}.
	\end{aligned}
	\label{eq:ris_min_elements}
\end{equation}
Using the simulation parameters and a representative SNR threshold of $20$ dB yields $M_{\min}=31$. Therefore, we set the values of $M$ as 16, 25, 36, 49, and 64 in our simulations. As shown in Fig.~\ref{RIS number}, the secrecy rates of all schemes increase with the RIS size due to enhanced passive beamforming gains. The proposed algorithm achieves the best performance across all RIS element quantities. By contrast, PDDGP updates optimization variables via dual-domain penalty and gradient projection operations. Although it can generate computationally efficient feasible solutions, its optimization capacity is inferior to the block-wise alternating optimization adopted in our framework. Shared coefficients and Random coefficients fail to fully exploit the spatial degrees of freedom offered by the RIS, thus yielding degraded performance. Specifically, Without AN lacks controllable artificial interference to mitigate information leakage. Meanwhile, BS ZF merely performs interference nulling towards users, which results in substantial information leakage in the directions of eavesdroppers and further leads to severe performance degradation for this baseline.

\begin{figure}
	\centering
	\includegraphics[width=\columnwidth]{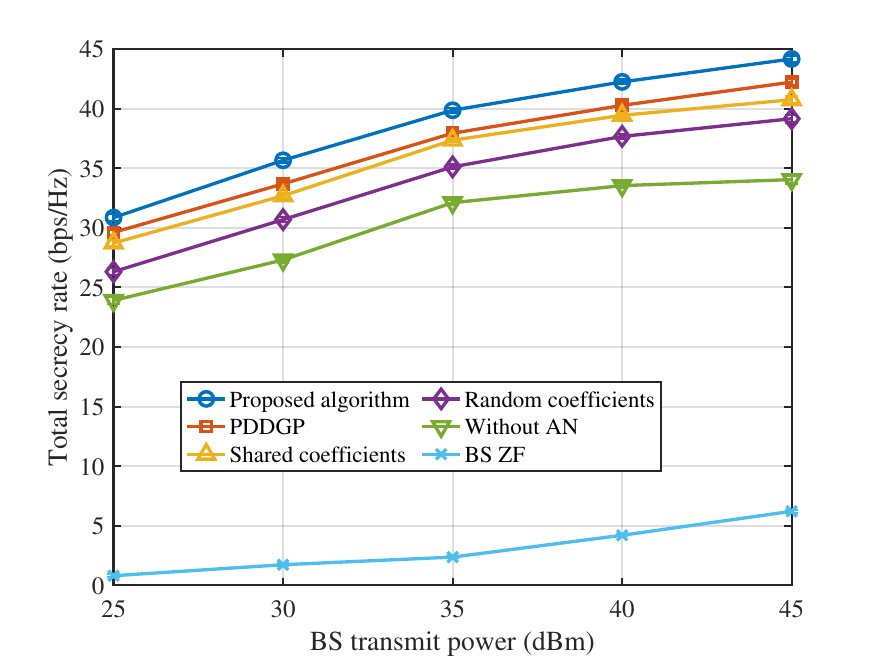}
	\caption{Total secrecy rate versus the BS transmit power.}
	\label{BS power}
\end{figure}

The impact of the BS transmit power budget on the achievable secrecy rate is examined in Fig.~\ref{BS power}. It can be observed that a higher BS power budget brings greater flexibility for beamforming and AN design, and accordingly improves the secrecy rate. Nevertheless, the growth of secrecy rate cannot be solely attributed to the increase of BS transmit power. Without proper power allocation, elevated transmit power will simultaneously strengthen the signal over the wiretap channel. By contrast, the proposed algorithm can efficiently convert the extra transmit power into performance gains to boost the total secrecy rate.

\begin{figure}
	\centering
	\includegraphics[width=\columnwidth]{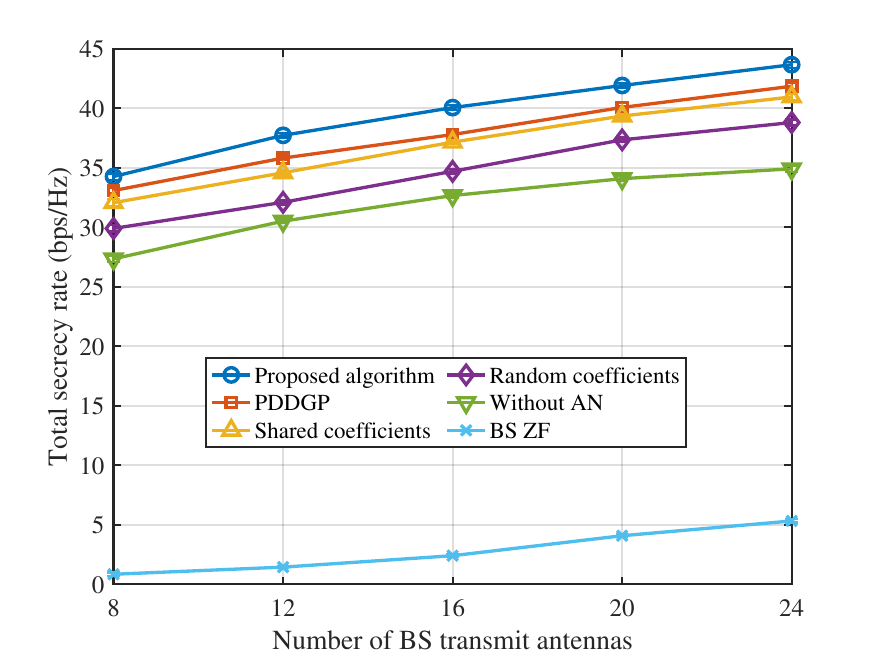}
	\caption{Total secrecy rate versus the number of BS transmit antennas.}
	\label{BS antenna}
\end{figure}

Fig.~\ref{BS antenna} plots the total secrecy rate as a function of the number of BS transmit antennas. Equipped with more transmit antennas, the BS acquires extra spatial degrees of freedom for transmit beamforming. This allows the BS to strengthen the desired signal at each user and curb information leakage toward the eavesdropper simultaneously, thereby boosting the total secrecy rate. Unlike raising the transmit power budget, increasing the number of transmit antennas expands the active spatial signal domain of the BS, which is particularly beneficial for serving multiple users and suppressing wiretap channels. As the antenna scale grows, the BS can better separate multiple target information beams, mitigate inter-user interference, and allocate AN toward directions that impose minimal performance degradation on users. Consequently, the performance superiority of the proposed algorithm becomes more pronounced with larger antenna arrays.

\begin{figure}
	\centering
	\includegraphics[width=\columnwidth]{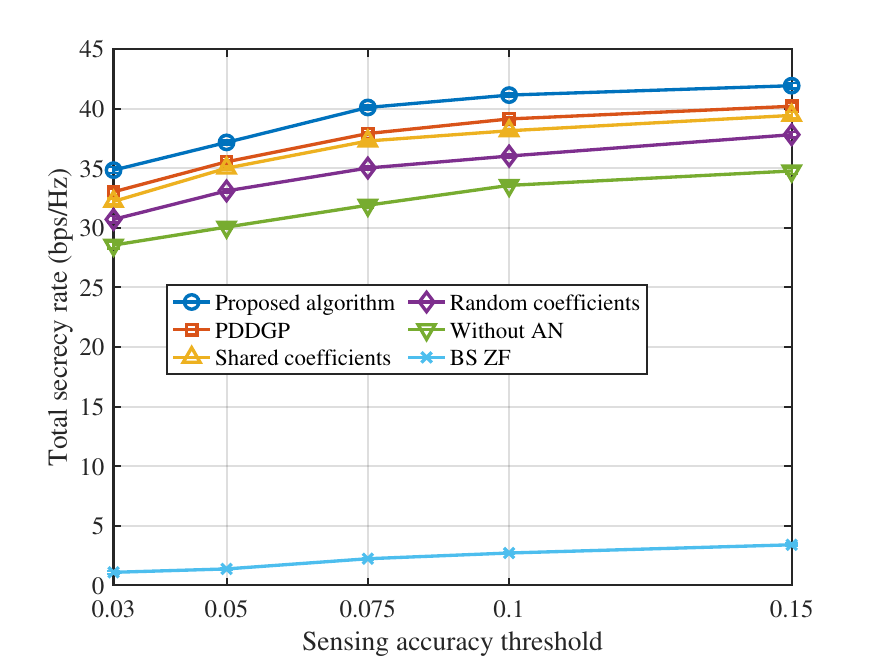}
	\caption{Total secrecy rate versus the tracking accuracy threshold.}
	\label{track}
\end{figure}

Fig.~\ref{track} illustrates the total secrecy rate versus the tracking accuracy threshold $\Gamma_{\max}$. It can be observed that the secrecy rate increases with $\Gamma_{\max}$, since relaxing the EKF tracking constraint enlarges the feasible region of the joint optimization problem. Consequently, greater flexibility is available for optimizing the BS beamforming, AN, and RIS reflection coefficients to enhance legitimate communication and suppress information leakage. In contrast, a stringent tracking constraint requires the system to maintain higher sensing accuracy, thereby limiting the degrees of freedom available for secrecy rate maximization. Moreover, the secrecy rate improvement becomes less pronounced as $\Gamma_{\max}$ increases, indicating diminishing gains from further relaxation of the tracking constraint. This suggests that the tracking accuracy threshold should be selected according to the application-specific estimation requirements, while considering the sensitivity of the secrecy rate to $\Gamma_{\max}$. In particular, a threshold near the onset of diminishing secrecy gains, provided that it satisfies the required tracking accuracy, can achieve a favorable balance between secrecy performance and tracking accuracy.

\begin{figure}
	\centering
	\includegraphics[width=\columnwidth]{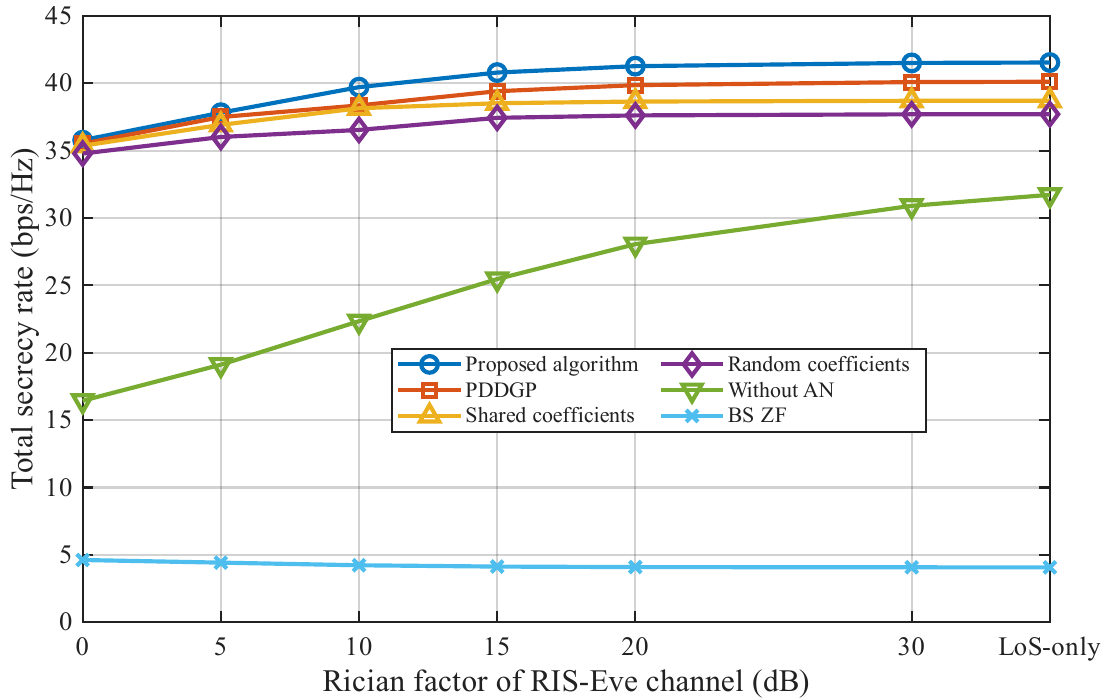}
	\caption{Total secrecy rate versus the Rician factor of the RIS-eavesdropper channel.}
	\label{rician}
\end{figure}

We further investigate the impact of RIS-eavesdropper channel mismatch on the secrecy performance, as shown in Fig.~\ref{rician}. Specifically, the proposed algorithm is optimized based on the LoS channel reconstructed from the EKF-predicted state, while the actual secrecy rate is evaluated under Rician fading with unknown NLoS components. It can be observed that the secrecy rate decreases as the Rician factor decreases, since stronger NLoS components introduce larger channel mismatch and degrade the effectiveness of secure transmission. Nevertheless, the proposed algorithm maintains superior secrecy performance compared with the benchmark schemes. As the Rician factor increases, the secrecy rate gradually approaches that of the LoS-only case, indicating that the proposed design remains effective in LoS-dominated environments despite channel mismatch.

\section{Conclusion}\label{wubufen}

This paper studied the total secrecy rate maximization problem for a RIS-enabled secure ISAC system with the mobile eavesdropper. To handle the time-varying eavesdropping channel, the continuous dynamic process was divided into multiple time slots, and the EKF was adopted to update the eavesdropper state. Based on the predicted state information, the BS beamforming, AN, and the RIS reflection coefficients for secure transmission and echo sensing were jointly optimized under the tracking accuracy constraint. To solve the resulting non-convex problem, an alternating optimization algorithm was developed. Specifically, the original problem was transformed by introducing auxiliary variables and decomposed into several tractable subproblems. The BS beamforming and AN, the RIS coefficients for secure transmission, the RIS coefficients for echo sensing, and the eavesdropping-rate upper bounds were updated alternately until convergence. Numerical results demonstrated that the proposed algorithm outperforms the benchmark schemes in terms of secrecy rate while effectively tracking the moving eavesdropper.

\begin{appendices} 
	\section{Derivation of the function $F$}

In this section, we explain the transformation process of the function $F$ in detail.
\begin{align}
	&\sum\limits_{k = 1}^K
	2\sqrt {1 + {r_k}}
	{\rm{Re}}\!\left\{
	c_k^*{{\bf{H}}_k}\!\left[n\right]
	{{\bf{w}}_k}\!\left[n\right]
	\right\}
	=
	{\rm{Re}}\!\left\{
	{{{\bf{a}}^{\rm{H}}}{\bf{w}}\left[n\right]}
	\right\},
	\\
	&\sum\limits_{k = 1}^K
	{{\left|c_k\right|}^2}
	\sum\limits_{c = 1}^K
	{{\left|
			{{\bf{H}}_k}\left[n\right]
			{{\bf{w}}_c}\left[n\right]
			\right|}^2}
	=
	{\left\|
		{{\bf{B}}_1}{\bf{w}}\left[n\right]
		\right\|^2},
	\\
	&\sum\limits_{k = 1}^K
	{{\left|c_k\right|}^2}
	{{\left|
			{{\bf{H}}_k}\left[n\right]
			{\bf{z}}\left[n\right]
			\right|}^2}
	=
	{\left\|
		{{\bf{B}}_2}{\bf{z}}\left[n\right]
		\right\|^2},
	\\
	&{\varepsilon _1}
	=
	\sum\limits_{k = 1}^K
	\left[
	{{\log}_2}\left(1+{r_k}\right)
	- r_k
	-{\left|c_k\right|^2}\sigma_k^2
	-A_k
	\right].
\end{align}
where
${\bf{a}}
=
\left[
{{\bf{a}}_1^{\rm T}},
{{\bf{a}}_2^{\rm T}},
\ldots,
{{\bf{a}}_K^{\rm T}}
\right]^{\rm T}$,
${{\bf{a}}_k}
=
2\sqrt{1+r_k}
c_k
{\bf{H}}_k^{\rm H}\left[n\right]$,
$
{{\bf{B}}_1}^{\rm H}{{\bf{B}}_1}
=
{{\bf{I}}_K}
\otimes
\left(
\sum\nolimits_{k = 1}^K
{\left|c_k\right|^2}
{\bf{H}}_k^{\rm H}\left[n\right]
{{\bf{H}}_k}\left[n\right]
\right),$
and
$
{\bf{B}}_2^{\rm H}{{\bf{B}}_2}
=
\sum\nolimits_{k = 1}^K
{\left|c_k\right|^2}
{\bf{H}}_k^{\rm H}\left[n\right]
{{\bf{H}}_k}\left[n\right].
$

Due to
${{\bf{H}}_k}\left[n\right]{{\bf{w}}_k}\left[n\right]
=
{\bf{h}}_k^{\rm H}
{\rm{diag}}\left(
{\bf{H}}_{\rm BR}{{\bf{w}}_k}\left[n\right]
\right)
{\bf{f}}\left[n\right]$
and
${{\bf{H}}_k}\left[n\right]{\bf{z}}\left[n\right]
=
{\bf{h}}_k^{\rm H}
{\rm{diag}}\left(
{\bf{H}}_{\rm BR}{\bf{z}}\left[n\right]
\right)
{\bf{f}}\left[n\right]$,
we have
\begin{align}
	&\sum\limits_{k = 1}^K
	2\sqrt {1+{r_k}}
	{\mathop{\rm Re}\nolimits}
	\left\{
	c_k^*
	{{\bf{H}}_k}\left[n\right]
	{{\bf{w}}_k}\left[n\right]
	\right\}
	=
	{\mathop{\rm Re}\nolimits}
	\left\{
	{{{\bf{g}}^{\rm H}}{\bf{f}}\left[n\right]}
	\right\},
	\\
	&\sum\limits_{k = 1}^K
	{{\left|c_k\right|}^2}
	\sum\limits_{c = 1}^K
	{{\left|
			{{\bf{H}}_k}\left[n\right]
			{{\bf{w}}_c}\left[n\right]
			\right|}^2}
	=
	{{\bf{f}}^{\rm H}}\left[n\right]
	{{\bf{D}}_1}
	{\bf{f}}\left[n\right],
	\\
	&\sum\limits_{k = 1}^K
	{{\left|c_k\right|}^2}
	{{\left|
			{{\bf{H}}_k}\left[n\right]
			{\bf{z}}\left[n\right]
			\right|}^2}
	=
	{{\bf{f}}^{\rm H}}\left[n\right]
	{{\bf{D}}_2}
	{\bf{f}}\left[n\right],
\end{align}
where
$\mathbf g^{\rm H}
=
\sum_{k=1}^{K}
2\sqrt{1+r_k}
c_k^*
\mathbf h_k^{\rm H}
\operatorname{diag}
\left(
\mathbf H_{\rm BR}\mathbf w_k[n]
\right)$,
$\mathbf D_1
=
\sum_{k=1}^{K}
|c_k|^2
\sum_{c=1}^{K}
\left(
\mathbf h_k^{\rm H}
\operatorname{diag}
\left(
\mathbf H_{\rm BR}\mathbf w_c[n]
\right)
\right)^{\rm H}
(
\mathbf h_k^{\rm H}
\operatorname{diag}
(
\mathbf H_{\rm BR} \\ \mathbf w_c[n]
)
)$,
and
$\mathbf D_2
=
\sum_{k=1}^{K}
|c_k|^2
\left(
\mathbf h_k^{\rm H}
\operatorname{diag}
\left(
\mathbf H_{\rm BR}\mathbf z[n]
\right)
\right)^{\rm H}
(
\mathbf h_k^{\rm H}\\
\operatorname{diag}
\left(
\mathbf H_{\rm BR}\mathbf z[n]
\right)
)$.
Let $\mathbf D=\mathbf D_1+\mathbf D_2$, thus we can derive
\begin{align}
	F
	&=
	{\mathop{\rm Re}\nolimits}
	\left\{
	{{{\bf{g}}^{\rm H}}{\bf{f}}\left[n\right]}
	\right\}
	-
	{{\bf{f}}^{\rm H}}\left[n\right]
	{\bf{D}}
	{\bf{f}}\left[n\right]
	+
	{\varepsilon _1},
	\nonumber\\
	&=
	{\mathop{\rm Re}\nolimits}
	\left\{
	{{{\bf{a}}^{\rm H}}{\bf{w}}\left[n\right]}
	\right\}
	-
	{\left\|
		{{\bf{B}}_1}{\bf{w}}\left[n\right]
		\right\|^2}
	-
	{\left\|
		{{\bf{B}}_2}{\bf{z}}\left[n\right]
		\right\|^2}
	+
	{\varepsilon _1}.
	\nonumber
\end{align}
\end{appendices}

\balance
\bibliographystyle{IEEEtran}
\bibliography{ref}

\end{document}